\documentclass{pasj01}

\Received{2020/05/15}
\Accepted{2020/07/07}
 
\begin{document} 

\title{ 
Supersoft X-Ray Phases of Recurrent Novae as an Indicator of 
their White Dwarf Masses
}

\author{Mariko \textsc{Kato}\altaffilmark{1}}
\altaffiltext{1}{Keio University, Hiyoshi, Yokohama
  223-8521, Japan}
\email{mariko.kato@hc.st.keio.ac.jp}

\author{Izumi \textsc{Hachisu}\altaffilmark{2}}
\altaffiltext{2}{Department of Earth Science and Astronomy, College of Arts and
Sciences, The University of Tokyo, 3-8-1 Komaba, Meguro-ku, Tokyo 153-8902, Japan}


\maketitle

\begin{abstract}
We have examined the optical/X-ray light curves of seven well-observed 
recurrent novae, V745~Sco, M31N 2008-12a, LMC~N 1968, 
U~Sco, RS~Oph, LMC~N 2009a, 
T~Pyx, and one recurrent nova candidate LMC~N 2012a. 
Six novae out of the eight show a simple relation that the duration of
supersoft X-ray source (SSS) phase is 0.70 times the total duration 
of the outburst 
($=$ X-ray turnoff time), i.e., $t_{\rm SSS}=0.70~t_{\rm off}$, 
the total duration of which ranges from 10 days to 260 days. 
These six recurrent novae show a broad rectangular X-ray light curve shape,
first half a period of which is highly variable in the X-ray count rate.
The SSS phase corresponds also to an optical plateau phase 
that indicates a large accretion disk irradiated by a hydrogen-burning WD.  
The other two recurrent novae, T~Pyx and V745~Sco, 
show a narrow triangular shape of X-ray light curve 
without an optical plateau phase.  Their relations 
between $t_{\rm SSS}$ and $t_{\rm off}$ are rather different from the 
above six recurrent novae. 
We also present theoretical SSS durations for recurrent novae
with various WD masses and stellar metallicities ($Z=$0.004, 0.01, 0.02,
and 0.05) and compare with observed durations of these recurrent novae.   
We show that the SSS duration is a good indicator of the WD mass 
in the recurrent novae with a broad rectangular X-ray light curve shape.  
\end{abstract} 


\section{Introduction} \label{introduction}

Novae occur in a binary of an accreting white dwarf (WD) and 
a non-degenerate companion. 
When the accreted H-rich envelope of the WD reaches a critical mass 
unstable H-burning sets in.
The H-rich envelope expands to a giant size and strong wind mass-loss occurs.
The winds are accelerated deep inside the photosphere at around the 
Fe peak of the radiative opacity ($\log ~T$ (K) $\sim 5.2$).
So they are called the optically thick winds. 
The wind mass loss continues during the expanded stage of the envelope, 
through the maximum expansion and in the optical decay phase. 
The winds stop after a large part of the envelope mass is blown in the wind. 
Then, the photosphere shrinks to a white dwarf size 
and the nova enters a supersoft X-ray source (SSS) phase.
The main emitting wavelength region of a nova shifts from optical 
to ultra-violet (UV), and finally to supersoft X-ray 
until hydrogen burning extinguishes.

The evolution timescales of optical and X-ray phases depend 
on the WD mass and chemical composition 
of the envelope. Based on the optically thick wind model \citep{kat94h}, 
\citet{hac06k} showed that the optical/infra-red (IR) 
light curves of classical novae are well reproduced with free-free 
emission from the wind, whereas the supersoft X-ray light curves 
are roughly reproduced with blackbody emission from the photosphere. 
\citet{hac06k} also found that all the optical/IR light curves decay as 
$F_\nu \propto t^{-1.75}$ independently of the WD mass and chemical 
composition.  Here, $F_\nu$ is the flux at the frequency
$\nu$ and $t$ is the time after the outburst.  They call these homologous
shapes of light curves ``the universal decline law.'' 
The universal decline law is well applied to
S (smooth) -type light curve shape novae defined by \citet{str10}.
The other types of nova light curve shapes deviate from the model
light curves in some part.  Such exceptions of nova light curves
include multi-peaks, secondary maximum, oscillations, dust blackout, 
jitters, plateau, shock contribution, and so on.  
However, their global trends of decline
can be sometimes fitted with the universal decline law.
From multiwavelength light curve fittings, they estimated the WD masses 
for many novae \citep{hac06k, hac07k, hac10k, hac15k, hac16ka, 
hac18a, hac18b, hac19a, hac19b}.
In general, a nova on a more massive WD evolves faster. 
Thus, the supersoft X-ray emergence time $t_{\rm on}$ 
(the end of the wind phase) and the supersoft X-ray turnoff time
$t_{\rm off}$ (the end of shell flash) has a positive 
linear correlation in the $(\log ~t_{\rm on})$-$(\log ~t_{\rm off})$
diagram (e.g. \citet{hac10k}).

A large sample of classical novae appeared in M31 has been presented by   
\citet{hen11} and \citet{hen14}. 
They reported statistical relation between various nova properties, 
such as the X-ray $t_{\rm on}$ and $t_{\rm off}$, X-ray blackbody temperature, 
and nova speed class. 
Figure 8(a) of \citet{hen14} shows the $t_{\rm on}$ versus
 $t_{\rm off}$ diagram,
in which the distribution of M31 classical novae indicates a positive 
correlation.
They made a power law fit as 
\begin{equation}
t_{\rm off}=1.0^{0.9 \pm1} \times t_{\rm on}^{0.8 \pm 1}. 
\label{equ_henze}
\end{equation}
This is different from the theoretical relation obtained 
by \citet{hac10k}, 
although Hachisu \& Kato's relation is located
in the middle of the scattered data of M31 novae. 
The reason for the difference is not yet clarified.

A recurrent nova is a nova with multiple recorded outbursts.  
The optical flux of a recurrent nova generally decays very fast 
followed by a short SSS phase. 
\citet{hac18b} examined light curves of very fast novae and recurrent novae. 
They showed that the recurrent novae do not always follow the universal
decline law of $F_\nu \propto t^{-1.75}$, in contrast to classical novae. 
The authors clarified the reason why the $V$ light curves deviate 
from the $t^{-1.75}$ law:  (1) contamination with radiation from shock 
between the ejecta and circumstellar material (V407 Cyg, RS Oph, 
and LMC~N~2009a), and 
(2) a too short duration of the $F_\nu \propto t^{-1.75}$ decline phase
followed by a much steeper decline phase of $F_\nu \propto t^{-3.5}$
(LMC~N~2012a). 
These effects influence individual recurrent novae differently, 
which results in a different shape of decline from the $t^{-1.75}$ law. 
Thus, the decline rate of the $V$/IR light curves is not a good indicator
of the WD mass in recurrent novae unlike classical novae.

\citet{hen15} pointed out a resemblance of light curves of three recurrent
novae, M31N 2008-12a, RS Oph, and U Sco: their X-ray light curves overlap
well each other in the normalized timescale with $t=0$ at their optical peaks. 
In other words, the ratio of the X-ray turn-on time and turnoff time, 
$t_{\rm on}/t_{\rm off}$, is common among the three novae, 
independently of the different timescales. 
\citet{bod16} also showed a similarity in the X-ray increase/decrease 
times between LMC 2009a and KT Eri (see their Figure 22).

Such a proportionality suggests that the recurrent novae has a 
different $t_{\rm on}$-$t_{\rm off}$ relation from classical novae. 
In the $t_{\rm on}$ - $t_{\rm off}$ diagram the recurrent nova M31N 2008-12a, 
having very short $t_{\rm on}$ and $t_{\rm off}$, is located at
the shortest edge of, but slightly below,  this relation 
(e.g., Figure 6 in \citet{hen15}). 
It is not clear if this excursion represents different physical property 
of recurrent novae from classical novae.

In the present work, we examine well-observed recurrent novae 
to see if they have a clear statistical relation similar to,
or different from, those of classical novae.
We further examine if the $t_{\rm on}$ - $t_{\rm off}$ relation is 
closely related to 
the WD mass, then the SSS duration could be an indicator of the WD mass.

This paper is organized as follows. 
In Section \ref{sec_sss}, we compare the $V$ and X-ray light curves 
of eight recurrent novae including a recurrent nova candidate 
and measure the X-ray turn-on/turnoff times. 
Then, we derive a scaling law between $t_{\rm on}$ and $t_{\rm off}$. 
Section \ref{sec_longsss} presents our theoretical calculation of 
SSS durations for recurrent novae on various WD masses and chemical
compositions. 
We then compare our calculated SSS durations with those of the recurrent
novae to examine if the SSS duration is a good indicator of the WD mass. 
Discussion and conclusions follow in Sections \ref{sec_discussion} 
and \ref{sec_conclusion}, respectively.


\begin{figure*}
 \begin{center}
\includegraphics[width=14cm]{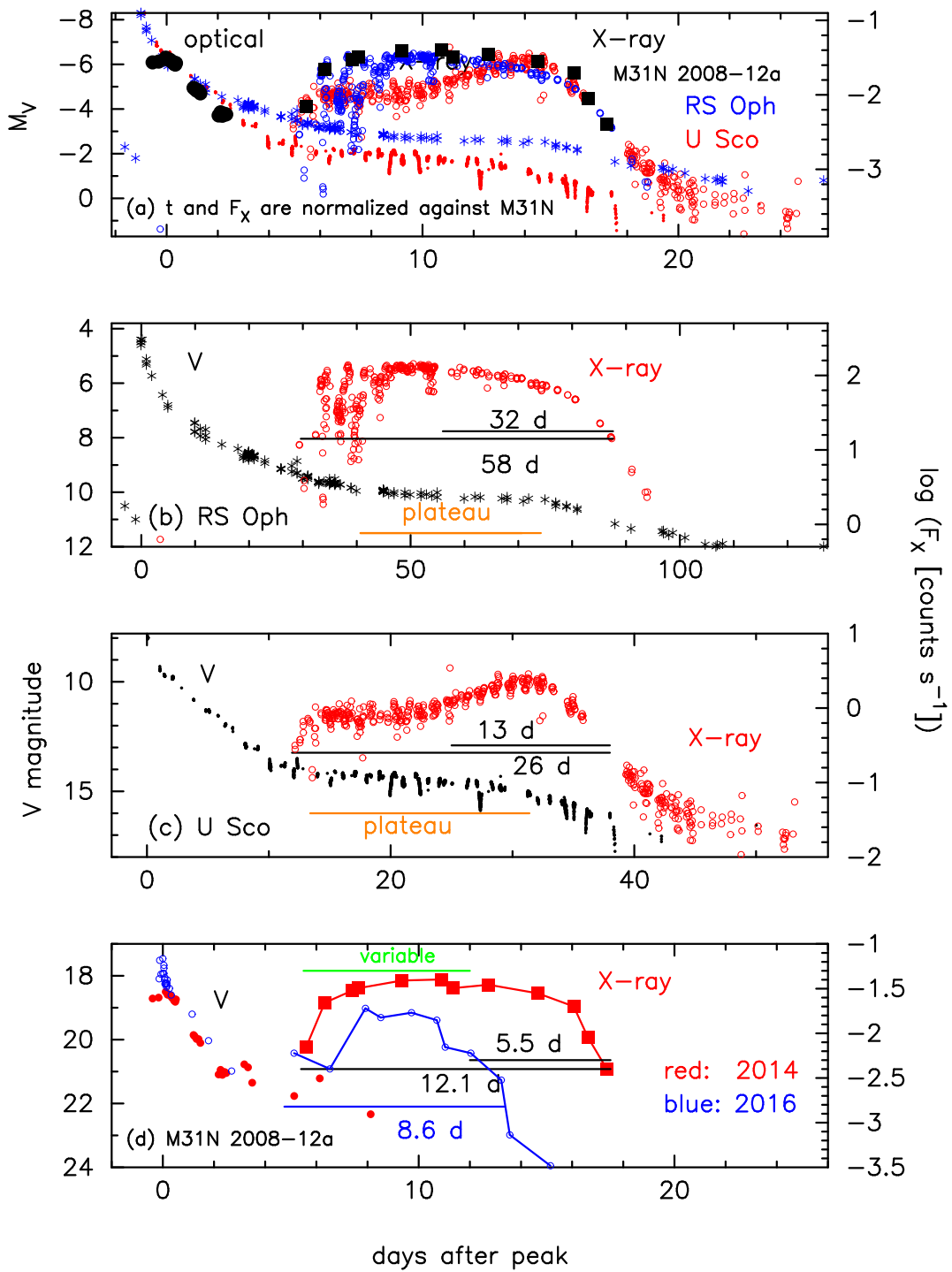}
 \end{center}
\caption{
Comparison of $V$ and X-ray light curves among the three recurrent novae, 
M31N 2008-12a, U Sco, and RS Oph. 
(a) Comparison in normalized timescale against the M31N 2008-12a 2014 
outburst. (b) RS Oph. (c) U Sco. (d) M31N 2008-12a in the 2014 (red symbols) 
and 2016 (blue) outburst. 
The horizontal black line segments indicate the duration of SSS phase, 
$t_{\rm SSS}$ (long segment) and $t_{\rm SSS}^*$ (short segment). 
The optical plateau phase is depicted by the horizontal orange line 
segment in panels (b) and (c).  The blue horizontal line segment in
panel (d) shows the highly variable phase in the 2016 outburst
taken from \citet{hen15, hen18}.  See text for details.
}\label{3rn}
\end{figure*}

\section{Scaling law of SSS phases in recurrent novae} 
\label{sec_sss}

Figure 7 of \citet{hen15} shows that the optical light curves of 
three recurrent novae (M31N 2008-12a, RS Oph and U Sco) are very similar 
in a normalized timescale and the X-ray light curves overlap well 
each other.  This means that the X-ray turn-on time and 
turnoff time have the same proportionality relation in these recurrent novae.
Figure \ref{3rn}(a) shows the X-ray and optical light curves 
of the same three recurrent novae as in \citet{hen15}, but for the absolute 
$V$ magnitude, $M_V$, of the optical light curves. 
Here, $M_V$ is obtained with the distance modulus in $V$ band,
$\mu_V\equiv (m -M)_V=$ 24.8 for 
M31N 2008-12a, 16.3 for U Sco, and 12.8 for RS Oph \citep{hac18b}. 
The X-ray count rates of U Sco and RS Oph are shifted in the vertical 
direction to fit with those of M31N 2008-12a. 
The light curves are squeezed in the time direction by a factor of 0.21
for RS Oph, and 0.48 for U Sco, against that of M31N 2008-12a. 

Figure \ref{3rn}(a) shows that the optical light curve of RS Oph decays
slightly slower than the other two novae, M31N 2008-12a and U Sco. 
Unlike many classical novae, recurrent novae do sometimes not follow 
the universal decline law, $F_\nu \propto t^{-1.75}$, as shown in Figure 24
of \citet{hac18b} and Figure 124 of \citet{hac19b}.  \citet{hac18b} and
\citet{hac19b} discussed the physical reason of the deviations.
RS Oph decays slowly like $F_\nu \propto t^{-1.55}$ 
because the $V$ flux is contaminated with emission from the shock
between the ejecta and circumstellar matter (CSM).   V407 Cyg shows 
a decay trend of $F_\nu \propto t^{-1.0}$ in the $V$/$I$ light curves 
because of much stronger shock heating than that of RS~Oph.  
U Sco follows a trend of $F_\nu \propto t^{-1.75}$ law in the very early
phase but soon transfers to a much steeper decline of 
$F_\nu \propto t^{-3.5}$ when the wind mass-loss rate sharply drops
before it enters the plateau phase owing to the disk irradiation. 
In M31N 2008-12a, the observational data are obtained 
only in the early phase and show a similar decay to U Sco. 
In this way, the differences in the optical light curves are
already discussed in \citet{hac18b} and \citet{hac19b}, 
so in the following subsections, we focus on the X-ray light curves,
especially their turn-on and turnoff times. 

\subsection{Definition of $t=0$, $t_{\rm on}$, and $t_{\rm off}$} 
In the present work we define the origin of time $t=0$ to be 
the optical peak. The X-ray turn-on time (turnoff time) is the time 
in units of days from $t=0$ when the X-ray flux rises (decreases) 
to a tenth of its peak value. 
Note that these definitions are different from those often used 
in X-ray observations of novae.

\subsection{RS Oph}\label{sec_rsoph}
Figure \ref{3rn}(b) shows the optical and X-ray light curves 
of RS Oph in the 2006 outburst. 
The optical data are the same as those in \citet{hac06b}, 
and the {\it Swift} X-ray data are taken from \citet{hac07kl}. 
\citet{hac06b} modeled the optical light curve with a combination 
of the WD photosphere, accretion disk, and companion star, both 
irradiated by the H-burning WD. 
They explained the plateau phase by the bright accretion disk 
irradiated by the WD, and concluded that the end of the plateau phase
corresponds to the end of hydrogen burning.

To compare the SSS duration with other recurrent novae, 
we define the SSS phase by the count rate larger than one tenth of
its peak value.  This phase is indicated by the longer 
black line segment in Figure \ref{3rn}(b). 
We obtain $t_{\rm on}= 30$ day, $t_{\rm off}= 88$ day, 
and the duration of $t_{\rm SSS}\equiv t_{\rm off}- t_{\rm on}= 58$ days.

The supersoft X-ray count rate varies two orders of magnitude 
in the early SSS phase until day $\sim 46$. 
This extremely variable phase was discussed in detail 
by \citet{osb11}. 
From the theoretical point of view, 
the optically thick wind becomes very weak 
just before its final stop in the early stage of the SSS phase. 
The envelope becomes optically thin as the mass-loss rate decreases. 
If the wind acceleration becomes unstable and correspondingly 
the wind mass-loss rate becomes variable, 
the supersoft X-ray flux could largely change from time to time.
Such a mechanism has not been studied yet, but it could be one of 
the explanations on a highly variable X-ray phase. 

If it is the case, we should allocate the highly variable X-ray phase
to still the theoretical wind phase, and regard the end of the variable
phase as the X-ray turn-on time.   Thus, the X-ray turn-on time becomes
longer while the SSS phase becomes shorter.  
We express them by $t_{\rm on}^*$  and $t_{\rm SSS}^*$ and distinguish them
from $t_{\rm on}$ and $t_{\rm SSS}$ which are defined by the period 
when the X-ray count rate is larger than one tenth of the maximum value.
As shown in Figure \ref{3rn}(b), 
the extremely variable phase ends on day 46 but, after that, the X-ray
flux seems to be still variable with small amplitudes.  Thus, we 
safely assume that the stable X-ray phase starts on  day 56.  
So we have $t_{\rm on}^*= 56$ day and $t_{\rm off}= 88$ day, and 
$t_{\rm SSS}^*= 32$ days.  This phase is depicted by the 
shorter black line segment, slightly above the longer black line segment,
in Figure \ref{3rn}(b).

\subsection{U Sco}\label{sec_usco}
Figure \ref{3rn}(c) shows the optical and X-ray (0.3 - 10 keV) light curves 
of U Sco in the 2010 outburst. The data are the same as those in 
\citet{hac18b}.  U Sco is a high inclination binary having the orbital 
period of 1.23 days \citep{schaefer10}.  
The narrow dips of the optical light curve 
in Figure \ref{3rn}(c) show eclipses by the companion star.  
The WD surface is occulted by the elevated disk surface, so we observe 
X-ray photons, not directly from the WD surface, 
but scattered by hot plasma around the WD via Thomson scattering 
\citep{nes12, ori13}. 
\citet{ori13} analyzed X-ray spectral lines on day 18 and 23 observed 
with {\it Chandra} and {\it XMM-Newton}, respectively, 
and concluded that mass-loss ceased between day 18 
and day 23 because the blue-shifted absorption component, 
owing to the WD corona, disappeared between them.  
The X-ray count rate also shows short-time variations which are attributed 
to the absorption by high-density clumps \citep{nes12,ori13}.  
\citet{ori13} concluded that the mass-loss was not a smooth process
but accompanied some clumpiness or inhomogeneity in the density around the WD.

Assuming a binary model consisting of an inflated WD, accretion disk
irradiated by the hot WD, and lobe-filling companion,
\citet{hac00a} and \citet{hac00b} reproduced the light curve of U Sco
1999 outburst.  The optical plateau phase is well explained by a large
contribution from the irradiated accretion disk.

We obtain $t_{\rm on}=12$ day, $t_{\rm off}=38$ day, 
and $t_{\rm SSS} =26$ days for the SSS phase of U~Sco
as indicated by the long black line segment in Figure \ref{3rn}(c). 
The X-ray variability seems to be large in the early phase and 
stabilized later. This variability may be related to partial obscuration 
by dense blobs falling around the WD or dense clumpy 
ejecta passing through the line of sight as discussed 
by \citet{nes12} and \citet{ori13}.  
If we exclude this variable X-ray phase, we have $t_{\rm on}^*= 25$ day, 
$t_{\rm off}= 38$ day, and $t_{\rm SSS}^*= 13$ days.

\subsection{M31N 2008-12a} \label{sec_m31n2008-12a}

M31N 2008-12a outbursted every year. 
The optical and X-ray light curves show remarkable resemblance 
every year except for the 2016 outburst \citep{dar16,hen18}. 
Figure \ref{3rn}(d) shows the X-ray and $V$ light curves of 
the 2014 outburst as a representative of the normal outburst. 
The data are taken from \citet{hen15} (X-ray: filled red squares connected 
by a red line) and \citet{dar15} (optical: red dots).  
The plateau phase is not clear in this year, but the 2015 data, 
which are not shown to avoid complication, 
clearly show the presence of a plateau phase \citep{dar16}. 
The X-ray variability of the supersoft X-ray phase was studied 
in detail by \citet{hen15} and by \citet{hen18}; the X-ray count rate shows 
large-amplitude and short-term variabilities 
until day 13 (in their definition) 
after the optical peak, but this variability suddenly
disappears after day 13.  Such a behavior has been reported 
in the all outbursts between 2013 and 2016.
 
For the 2014 outburst 
we obtain $t_{\rm on}=5.4$ day, $t_{\rm off}=17.5$ day, and  
$t_{\rm SSS} =12.1$ days for the full SSS duration of M31N~2008-12a
as indicated by the long black line, 
and $t_{\rm on}^*= 12.0$ day and $t_{\rm SSS}^*= 5.5$ days
by the short black line in Figure \ref{3rn}(d). 
 
Figure \ref{3rn}(d) also show the 2016 outburst in which the 
data are taken from \citet{hen18}.
It showed unusual behavior; the optical peak is one magnitude 
brighter, the X-ray count rate is 
smaller, and the SSS phase ended earlier. 
It is difficult to determine the X-ray turn-on time because
the X-ray data are not smooth. 
We adopt $t_{\rm on}=4.75$ day, $t_{\rm off}=13.35$ day, and
$t_{\rm SSS} =8.6$ days, assuming the X-ray flux rises just before 
the first data. However, the X-ray flux could increase earlier, 
say 0.5 days or so. The X-ray count rate also shows variability 
but with relatively small amplitudes, thus we do not define 
$t_{\rm on}^*$ and $t_{\rm SSS}^*$ for the 2016 outburst. 

\citet{hen18} presented an idea to explain the shorter SSS phase 
in the 2016 outburst. 
If the mass accretion resumes soon after the X-ray turn-on, 
and the WD accretes matter throughout the SSS phase, the duration of 
which becomes longer because fresh nuclear fuel is continuously 
supplied. On the other hand, if the mass-accretion does not restart soon 
because the accretion disk is blown by the wind, the SSS ends 
earlier than the continuous accretion case. 
In the 2016 outburst the brighter optical peak suggests that 
the outburst is stronger than the other years and, as a result,
the accretion disk could be completely destroyed. 
So the mass supply does not occur soon and the SSS duration is shorter. 
The unusual 2016 outburst indicates that the WD mass is not the 
only factor that determines the SSS duration in recurrent novae. 

\begin{figure*}
 \begin{center}
\includegraphics[width=14cm]{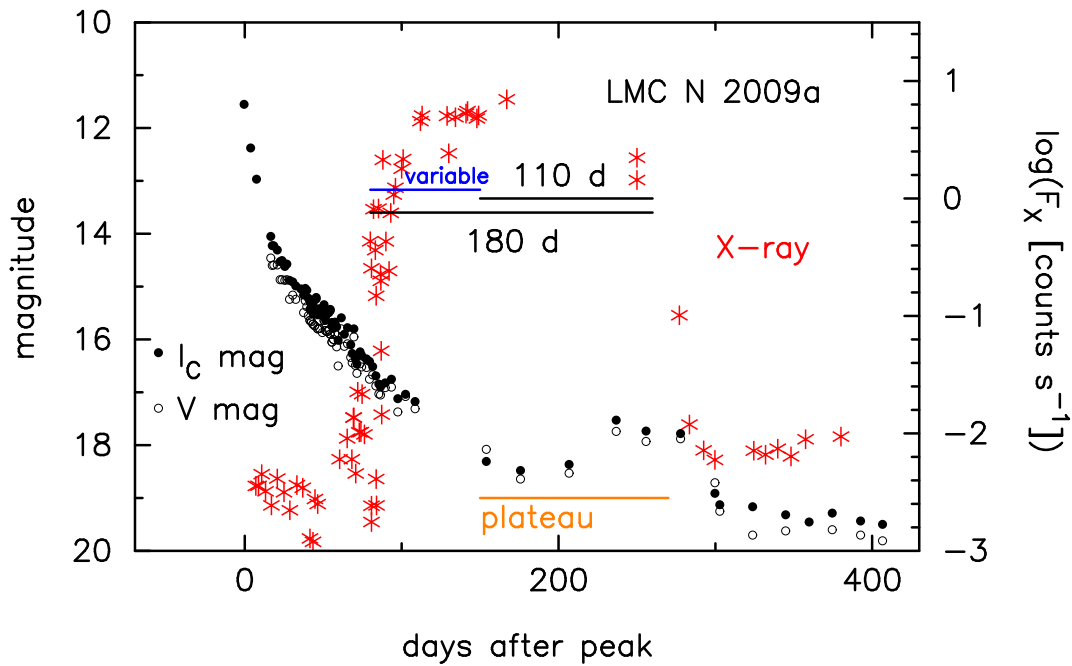}
 \end{center}
\caption{Same as Figure \ref{3rn}, but for LMC N 2009a.
The red asterisks indicate the X-ray count rate (0.3 - 10 keV) 
observed with {\it Swift}.  The filled black circles and 
open black circles show $I_{\rm C}$ and $V$ data, respectively,
taken from SMARTS \citep{wal12}. 
}\label{lmc2009}
\end{figure*}

\subsection{LMC~N~2009a} \label{sec_lmc2009a}

LMC nova 2009a is a recurrent nova with recorded outbursts in 1971 and 2009. 
Figure \ref{lmc2009} shows the X-ray, $V$ and $I$ light curves of 
LMC N 2009a in the 2009 outburst. 
The data are the same as those in Figure 33 of \citet{hac18b}. 

\citet{bod16} reported the pan-chromatic observations of the 2009 outburst
of LMC~N~2009a,  including UV and X-ray (0.3 - 10 keV) observed by
{\it Swift}, and $V$ and $I$ magnitudes observed with 
the Small and Medium Aperture Telescope System (SMARTS) \citep{wal12}.  
This nova is a short orbital period binary and its position in the 
$(B-R)_0$-$M_B$ color-magnitude diagram is close to U Sco, which indicates
the presence of a bright and hot accretion disk around the WD.   
Here, $(B-R)_0$ is the intrinsic $B-R$ color. 
The {\it Swift}/UVOT-uvw2 light curve shows a plateau phase 
until $\sim 280$ days after the discovery, indicating a UV-bright disk. 

\citet{bod16} also reported the strong variability in X-ray. 
Their Figure 11 shows that the X-ray flux varies largely 
in the very early SSS phase but becomes static in the later phase. 
The end of this variable SSS phase is not clear 
because the {\it Swift} observation mode changed from a high 
to low cadence.  We assume that the strongly variable phase 
ended on day 150 as indicated by the blue line 
in Figure \ref{lmc2009}. 
Thus, we have $t_{\rm on}= 80$ day, $t_{\rm off}= 260$ day, 
$t_{\rm SSS}= 180$ days, $t_{\rm on}^*= 150$ day, 
and $t_{\rm SSS}^*= 110$ days.

\citet{hac18b} analyzed the optical light curve of LMC~N~2009a in detail. 
They pointed out a resemblance of the shape of light curves 
between RS Oph and LMC N 2009a both for X-ray and $V$, 
even though the evolution timescale of RS Oph 
is 3.16 times longer (see their Figure 33). 
This means that LMC N 2009a also shows the same proportionality relation
as that of the three recurrent novae in Figure \ref{3rn}.

\begin{figure*}
 \begin{center}
\includegraphics[width=14cm]{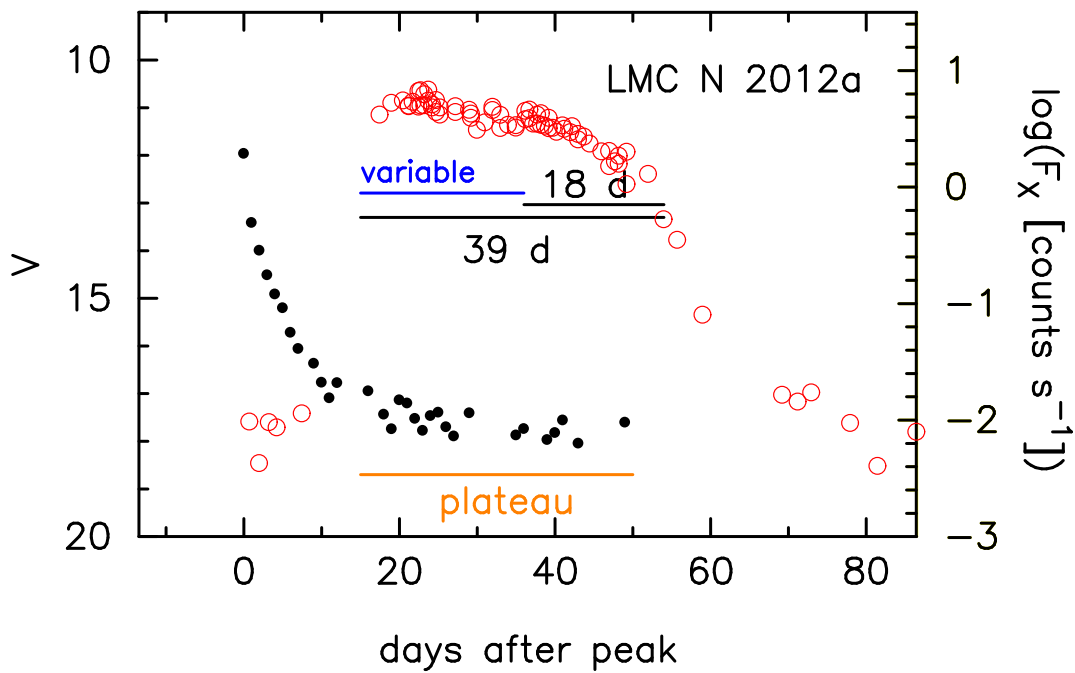}
 \end{center}
\caption{Same as Figure \ref{3rn}, but for LMC N 2012a. 
}\label{lmc2012}
\end{figure*}

\subsection{LMC~N~2012a: A recurrent nova candidate} 
\label{sec_lmc2012a}
\citet{hac18b} pointed out that the $B$, $V$ and $I_{\rm C}$ light curves
of LMC N 2012a overlap with those of U Sco including the optical plateau phase 
(see their Figures 26, 35 and 45). 
Moreover, the $(B-V)_0$ color of LMC N 2012a in the plateau phase 
agrees well with that of U Sco. 
Here, $(B-V)_0$ is the intrinsic $B-V$ color. 
Because the optical plateau phase of U Sco is dominated by the irradiated 
accretion disk, they concluded that LMC~N~2012a also has a bright 
irradiated accretion disk. 
These resemblance in the light curves and colors as well as a
relatively longer orbital period of 0.802 days \citep{schw15}
strongly suggests that LMC~N~2012a is a recurrent nova similar to U~Sco. 
\citet{hac18b} estimated the WD mass to be $1.37~M_\odot$ 
from a monotonic relation between $M_{\rm WD}$ and time-stretching factor 
(their Section 5.3). 

Figure \ref{lmc2012} shows the X-ray and $V$ light curves,
the data of which are taken from \citet{hac18b}.
We obtain $t_{\rm on}= 15$ day, $t_{\rm off}= 54$ day, and
$t_{\rm SSS}= 39$ days for the full SSS duration of LMC~N~2012a. 

\citet{schw15} reported the {\it Swift} and {\it Chandra} X-ray observations. 
The X-ray count rate is weakly variable until day $\sim 35$, followed by 
a smooth decline (see their Figure 14). 
The hardness ratio defined by ($0.5-10$ keV)$/$($0.3-0.5$ keV)
is also variable until day $\sim 37$ (their Figure 4). 
The exact time of switch from the variable to non-variable phase 
is not reported, so we define it as $t=36$ day 
as shown in Figure \ref{lmc2012}. 
We have $t_{\rm on}^*= 36$ day, $t_{\rm off}= 54$ day, 
and $t_{\rm SSS}^*= 18$ days.

\begin{figure*}
 \begin{center}
  \includegraphics[width=14cm]{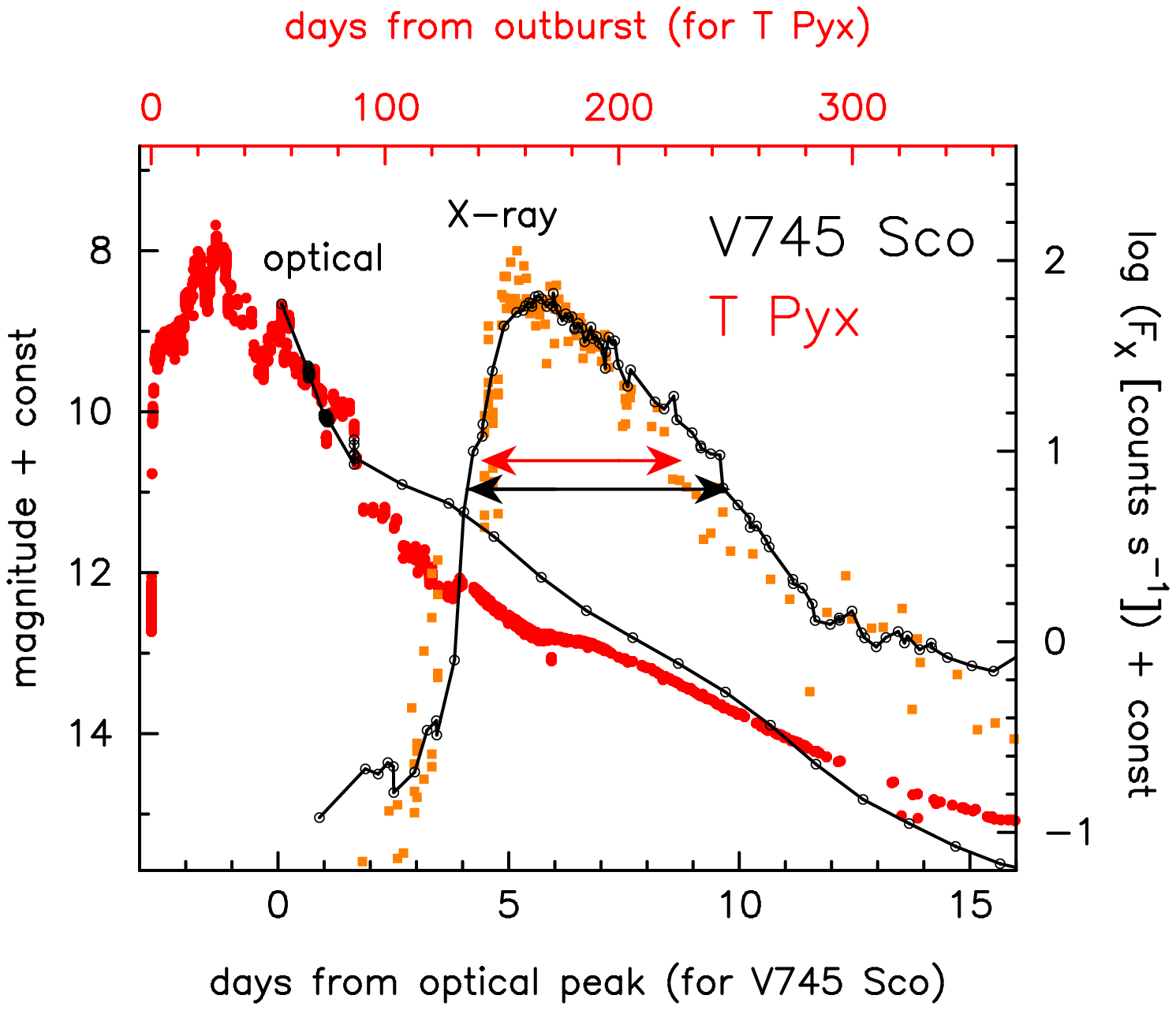}
 \end{center}
\caption{
The X-ray and optical light curves of the galactic recurrent novae, 
V745 Sco in the 2014 outburst (small black open circles connected 
by a black line: timescale is in the lower axis) and 
T Pyx in the 2011 outburst (the timescale in the upper side).
The optical $V$ light curve of T~Pyx is indicated by the red dots. 
Its magnitude is shifted downward by $1.5$ mag (ranges from $14.2$ mag
to $5.2$ mag).  The X-ray count rates of T~Pyx are indicated 
by the filled orange squares in the range from $-3.3$ to $0.5$.  
The X-ray duration is indicated to be $5.7$ days 
by the two-headed black arrow for V745 Sco, 
and to be $86$ days by the tow-headed red arrow for T Pyx. 
}\label{v745scotpyx}
\end{figure*}

\subsection{LMC N 1968}

LMC N 1968 outbursted in 1968, 1990, 2002, 2010, 2016, 
and 2020. \citet{kui20} presented detailed information on 
multiwavelength light curves of the 2016 outburst
including optical and X-ray bands. 
The optical/IR light curves show a plateau on day 10 - 25 corresponding 
to the SSS phase. The 2020 outburst data are not published yet, 
so we use the X-ray light curve in the 2016 outburst.  

Figure 8 in \citet{kui20} shows the X-ray count rate of LMC N 1968.  
There are no data in the middle 10 days. We assume the X-ray light 
curve has a flat peak and the flux keeps the same value during 
the missing period. The X-ray turn on/off time is not clear 
because the flux is not smooth but stays around one tenth of 
the assumed peak flux.  Under the condition we obtain the time since
the first discovery (2016 January 21.2094 UT) to be $t_{\rm on}= 7.2$ day,
$t_{\rm off}= 37.4$ day and $t_{\rm SSS}=30.2$ days.  There is no 
information on the presence or non-presence of strong variability. 

\subsection{V745 Sco}
V745 Sco is a symbiotic recurrent nova with recorded outbursts 
in 1937, 1989, and 2014.   The 2014 outburst of V745 Sco evolves
extremely fast such that the supersoft X-ray flux increased
only four days after the optical discovery. 
\citet{pag15} presented the multiwavelength light curves of IR, 
optical, UV and X-ray bands.  From the X-ray spectral analyses, 
the heavy element abundance of V745 Sco was estimated to be mildly
sub-solar, that is, Fe/Fe$_\odot=0.5$ and $0.3-0.7$ by \citet{ori15} and
\citet{dra16}, respectively.   
Figure \ref{v745scotpyx} shows the X-ray light curve obtained 
with {\it Swift} (0.3 - 2 keV:  taken from \citet{pag15}) as well as 
the $V$ light curve (taken from 
the American Association of Variable Star Observers (AAVSO)
until day 1.65, and SMARTS from day 1.69). 
The X-ray count rate goes up very quickly to reach the maximum, 
followed by a slow decay without a plateau phase.
No highly variable phase is detected. 
\citet{pag15} obtained the X-ray temperature from blackbody fit.
Their X-ray temperature increased and then decreased with the X-ray
count rate. 
They interpreted the X-ray decay phase to be the WD cooling phase.

Such a very fast evolution suggests that V745 Sco hosts 
an extremely massive WD. 
\citet{hac18b} fitted the X-ray light curve of V745 Sco with a  
theoretical model of 1.385 $M_\odot$ WD (see their Figure 7).
The envelope mass is very small and a substantial part of the ignition mass
is already blown in the wind before the X-ray flux turns on. 
This small envelope mass results in a quick rise and immediate drop
in the X-ray count rate, skipping a plateau phase of X-ray. 
In this model, the X-ray decay phase corresponds to the WD cooling phase,
being consistent with the interpretation by \citet{pag15}. 

The X-ray light curve of V745 Sco shows a narrow ``triangle'' shape,
not a broad ``rectangle'' shape like the other six novae.  
The X-ray plateau phase, like in RS Oph, corresponds to the steady
nuclear-burning phase \citep{hac06b}, where the envelope mass gradually
decreases owing to hydrogen burning. 
If the envelope mass is very small, the steady burning phase is also short.
V745 Sco has neither highly-variable X-ray phase nor flat peak. 
The envelope mass may be extremely small and mostly blown off 
before it enters an SSS phase. 
This is the reason for the narrow triangle shape of X-ray light curve. 
In the cooling phase of the WD, the X-ray flux drops quickly, 
so the irradiation effect of the disk is small. 
Even if V745 Sco has a large disk, we do not expect an optical
plateau phase in the cooling phase of the WD.
This is the reason why V745 Sco shows no plateau phase in the $V$ 
light curve.

Although the X-ray light curve shape is different, we adopt 
the same definition of X-ray turn on/off time, i.e., the epoch of 
one tenth of the peak X-ray flux as the other recurrent novae. 
This phase is indicated by the two-headed black arrow
in Figure \ref{v745scotpyx}. 
We obtain $t_{\rm on}= 4.1$ day, $t_{\rm off}= 9.8$ day and
$t_{\rm SSS}= 5.7$ days.

\subsection{T Pyx}

T Pyx is a recurrent nova with the recored outbursts in 1890, 1902,
1920, 1944, 1967 and 2011 \citep{schaefer10, sch13}.
The optical light curve is characterized 
by a relatively slow evolution and multiple peaks \citep{nel14,cho14}. 
Its orbital period, 0.076 days (=1.8 hr) \citep{sch13}, is located
below the period gap.  The orbital period suggests an exceptionally
low-mass companion, although very high mass-accretion
rates are usually not expected in such a short orbital period companion. 
The multiple peak in the optical light curve suggests that 
a single explosion model may not be applicable. 
Moreover, as shown later, the X-ray light curve has a narrow triangle 
shape without a broad flat peak.  
These properties make T Pyx very different from the other recurrent novae. 

The X-ray and optical light curves of the 2011 outburst are 
shown in Figure \ref{v745scotpyx}. 
The optical $V$ data (red dots) are taken from AAVSO. 
The X-ray count rates obtained with {\it Swift} (filled orange squares) 
are taken from the {\it Swift} website \citep{eva09}.
The X-ray light curve shows a quick rise followed by 
a slower decrease. Its shape is a triangle. 
We obtain $t_{\rm on}= 141$ day, $t_{\rm off}= 227$ day and
$t_{\rm SSS}= 86$ days.

Figure \ref{v745scotpyx} compares the light curve of T Pyx with that of
V745 Sco. The timescale ratio is 20. 
Here we fit the optical maximum of V745 Sco with the last 
optical maximum of T Pyx. 
Despite of a 20 times slower evolution in T~Pyx,
both the X-ray light curves show a homologous triangle shape. 
Because there are no theoretical light curve fitting models for T Pyx,
we do not go into detail of the evolution of this atypical recurrent nova.

\begin{figure*}
 \begin{center}
\includegraphics[width=14cm]{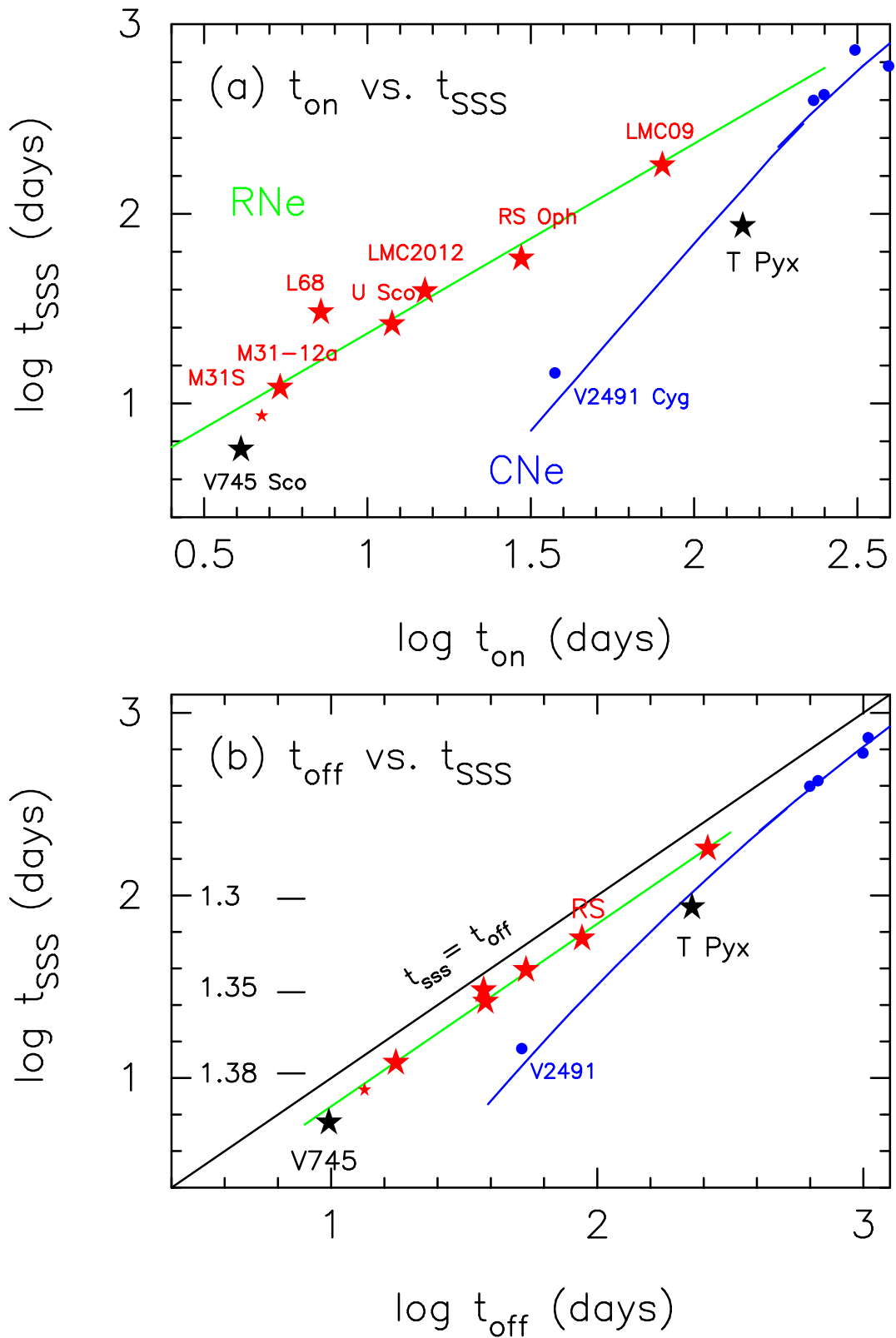}
 \end{center}
\caption{
The duration of the SSS phase of various recurrent and classical novae
against (a) X-ray turn-on time, and (b) X-ray turnoff time.
The red stars indicate, from left to right, M31N 2008-12a 
(small symbols for the 2016 outburst and 
large symbols for the 2014 outburst), 
LMC N 1968, U Sco, LMC N 2012, RS Oph, and LMC N 2009a. 
The black stars indicate the recurrent novae, 
V745 Sco (left) and T Pyx (right). 
The five blue dots represent the position of the classical novae, 
from lower left to upper right, V2491 Cyg, V1974 Cyg, V1494 Aql, 
V458 Vul, and V1213 Cen.
The blue line shows the theoretically derived relation for classical novae 
(see Appendix \ref{tsss-versus-toff}). 
The green lines represent the empirical relations, i.e.,
Equation (\ref{equ_sss}) in panel (a), 
but Equation (\ref{equ_sss2}) in panel (b). 
The black line in panel (b) indicates the relation, $t_{\rm SSS}=t_{\rm off}$.
In panel (b) LMC N 1968 is located very close to, but slightly above, U Sco.  
}\label{xonoff.nocorr.rn}
\end{figure*}

\subsection{Scaling Law of Recurrent Novae}

In the previous subsections, we have obtained the durations of SSS phases 
for eight recurrent novae including one candidate. 
Figure \ref{xonoff.nocorr.rn} shows $t_{\rm SSS}$ against (a) $t_{\rm on}$ 
and (b) $t_{\rm off}$ of these eight recurrent novae.
The five novae, M31N 2008-12a, U Sco, LMC N 2012a, RS Oph, and LMC N 2009a 
are located on a straight line. 
We draw a green straight line of inclination 1.0 
to fit these five nova data, which is defined by
\begin{equation}
\log t_{\rm SSS}=\log t_{\rm on}+0.37,
\label{equ_sss}
\end{equation}
in panel (a) and 
\begin{equation}
\log t_{\rm SSS}=\log t_{\rm off}-0.155,
\label{equ_sss2}
\end{equation}
in panel (b). 
From these two equations, we have the proportionality relations,
\begin{equation}
 t_{\rm on}=0.30 ~ t_{\rm off},  
\label{equation.on}
\end{equation}
and 
\begin{equation}
t_{\rm SSS}=0.70 ~ t_{\rm off}.
\label{equation.off}   
\end{equation} 
LMC N 1968 is located substantially above the green line of 
Equation (\ref{equ_sss}). 
We suppose that this deviation comes either from scatter
of the X-ray data points around $t_{\rm on}$ and $t_{\rm off}$ 
or from some physics that keeps the X-ray count rate high 
at one tenth of the peak level that is not observed
in the other recurrent novae.  In the 2020 outburst or after,
we expect that much more detailed information on the X-ray flux is available. 

The two recurrent novae with a triangle shape of X-ray light curve
are located much below the green line of Equation (\ref{equ_sss})
in Figure \ref{xonoff.nocorr.rn}(a).  The position of
T~Pyx is rather close to the theoretical line for classical novae (blue). 
V745 Sco is located near the shortest limit of the curving alignment 
of recurrent novae that seems to apart downward from the green line in panel (a).  
Note, however, that $t_{\rm on}$ of M31N 2008-12a in the 2016 outburst 
(small star) is very uncertain. 
The position of M31N 2008-12 2016 outburst could be in the
middle of M31N 2008-12a 2014 outburst and V745 Sco as in Figure \ref{xonoff.nocorr.rn}(a),
but it could move toward upper-left and approach 
closer to the green line, if we assume a shorter, say 0.5 days shorter 
$t_{\rm on}$. Thus, it is uncertain whether V745 Sco is the shortest edge 
of the curving alignment (sudden drop from the green line at M31N 2008-12a) 
of the recurrent novae, or an exception from the straight line, 
Equation (\ref{equ_sss}).  

To clarify the difference of recurrent novae from classical novae, 
we plot the theoretical relation 
derived for the classical nova models \citep{hac10k} 
(see Appendix \ref{tsss-versus-toff} for derivation). 
We also added five classical novae denoted by the blue dots. 
They are, from lower-left to upper-right, 
V2491 Cyg (data taken from Figure 1 in \citet{hac09}, \citet{pag10,nes11}), 
V1974 Cyg (Figure 38 in \citet{hac16ka}), 
V1494 Aql (Figure 22 in \citet{hac10k}), 
V458 Vul (Figure 67 in \citet{hac16c}), 
and V1213 Cen (Figure 65 in \citet{hac19b}). 
These five novae are located consistently on the blue line. 
The position of V2491 Cyg demonstrates that classical novae 
have different physical properties from recurrent novae. 
These two lines are close in the longer $t_{\rm on}$ (or $t_{\rm off}$), 
but largely different in the shorter $t_{\rm on}$ (or $t_{\rm off}$). 
We safely conclude that recurrent novae have a longer SSS 
phase than classical novae that have the same $t_{\rm on}$ 
(or $t_{\rm off}$) time.

\begin{figure*}
 \begin{center}
\includegraphics[width=14cm]{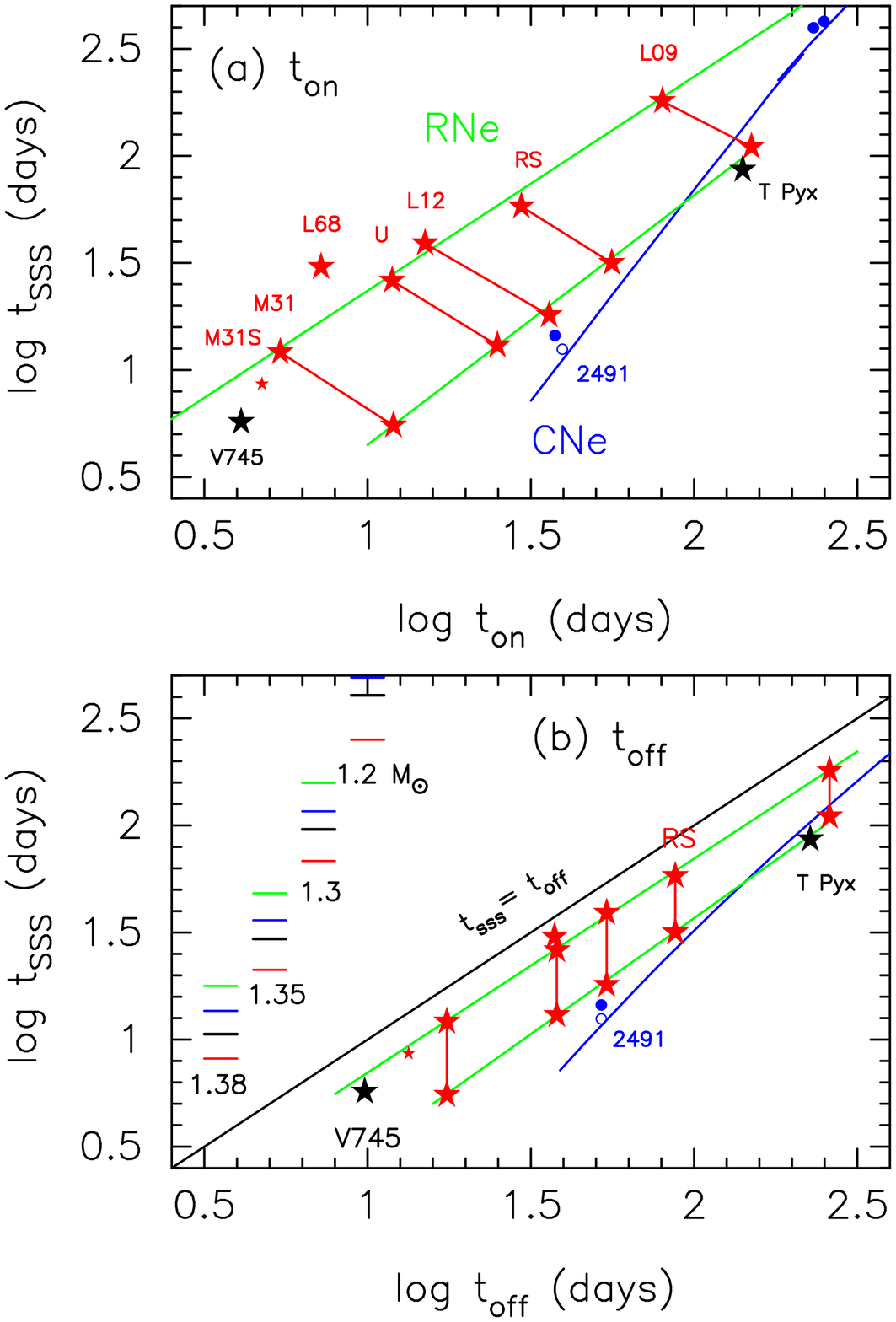}
 \end{center}
\caption{
Same as Figure \ref{xonoff.nocorr.rn}, but we added $t_{\rm SSS}^*$, 
the SSS duration excluding the highly variable phase, that are  
indicated by the lower red stars connected with each $t_{\rm SSS}$ point
(above red stars). 
The lower green straight lines indicate empirical relations,  
$t_{\rm SSS}^*$ vs. $t_{\rm on}^*$ of Equation (\ref{equ_sss3}) 
in panel (a), but $t_{\rm SSS}^*$ vs. $t_{\rm off}$ of
Equation (\ref{equ_sss4}) in panel (b).  See text for more details.  
}\label{xonoff.rn}
\end{figure*}

Figure \ref{xonoff.rn} shows the same $t_{\rm SSS}$ as those 
in Figure \ref{xonoff.nocorr.rn}, but we added another definition of
the SSS phase, which starts after the variable X-ray phase ends, 
$t_{\rm SSS}^*$.  
The five novae are located along the lower green line, which is defined by 
\begin{equation}
\log ~t_{\rm SSS}^*= 1.17 ~ (\log ~t_{\rm on}^*-1.0) + 0.65,
\label{equ_sss3}
\end{equation}
in Figure \ref{xonoff.rn}(a) and  
\begin{equation}
\log ~t_{\rm SSS}^*=1.08 ~ (\log ~t_{\rm off}-1.2)+ 0.7, 
\label{equ_sss4}
\end{equation}
in Figure \ref{xonoff.rn}(b). 
These lines are calculated from the positions of the five recurrent novae
with a broad rectangular SSS phase, and not parallel but slightly steeper
than Equations (\ref{equ_sss}) and (\ref{equ_sss2}). 
The two SSS periods, $t_{\rm SSS}$ and $t^*_{\rm SSS}$, 
of each nova are connected by the line.  
Interestingly, T~Pyx is located on the lines of $t_{\rm SSS}^*$
in both the upper and lower panels. 
Its position is also close to the blue classical nova line.  

The classical nova V2491 Cyg also shows highly variable SSS phase
as reported by \citet{nes11}. 
Their Figure 1 demonstrates a deep dip on day $\sim 39$ in the 
{\it Swift} X-ray count rate. 
If we exclude the part before the dip, we have a two-days shorter  
period, $t_{\rm SSS}^*= 12.5$ days compared with $t_{\rm SSS}= 14.5$ days 
(blue open circle in Figure \ref{xonoff.rn}).
In both the upper and lower panels, $t_{\rm SSS}^*$ of V2491 Cyg
is more closer to the theoretical line of classical novae (blue line).

\begin{figure}
  \begin{center}
  \includegraphics[width=8.cm]{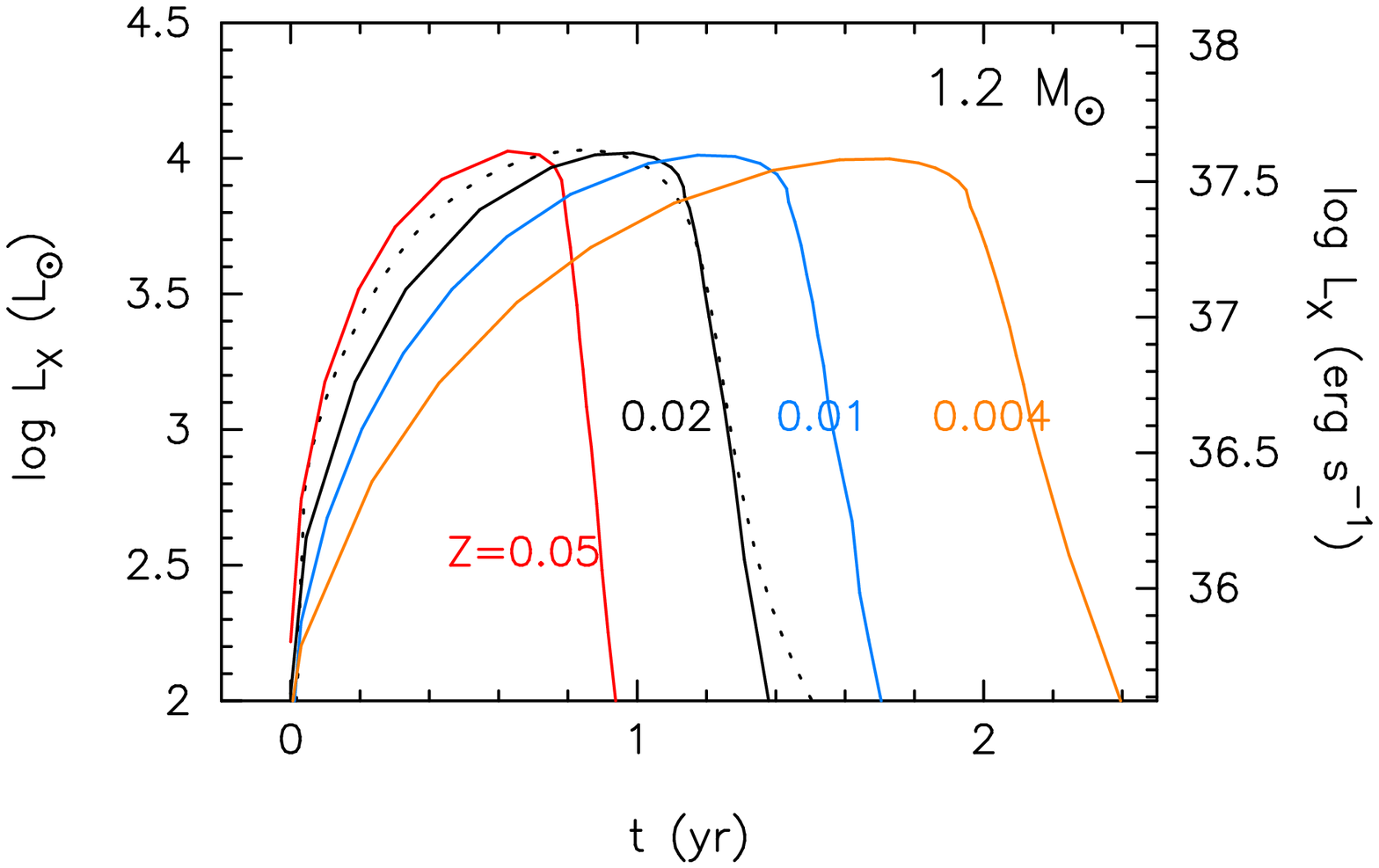}
 \end{center}
\caption{
The X-ray ($0.3- 1.0$ keV) light curve in the SSS phase 
of a recurrent nova on a 1.2 $M_\odot$ WD with the metallicity of
$Z=$0.004 (orange), 0.01 (blue), 0.02 (black), and 0.05 (red). 
The dotted line indicates the evolution model for a 1.2 $M_\odot$ WD 
with $Z=0.02$ \citep{kat17sha}. 
\label{sss12}}
\end{figure}

\begin{figure}
  \begin{center}
  \includegraphics[width=8.cm]{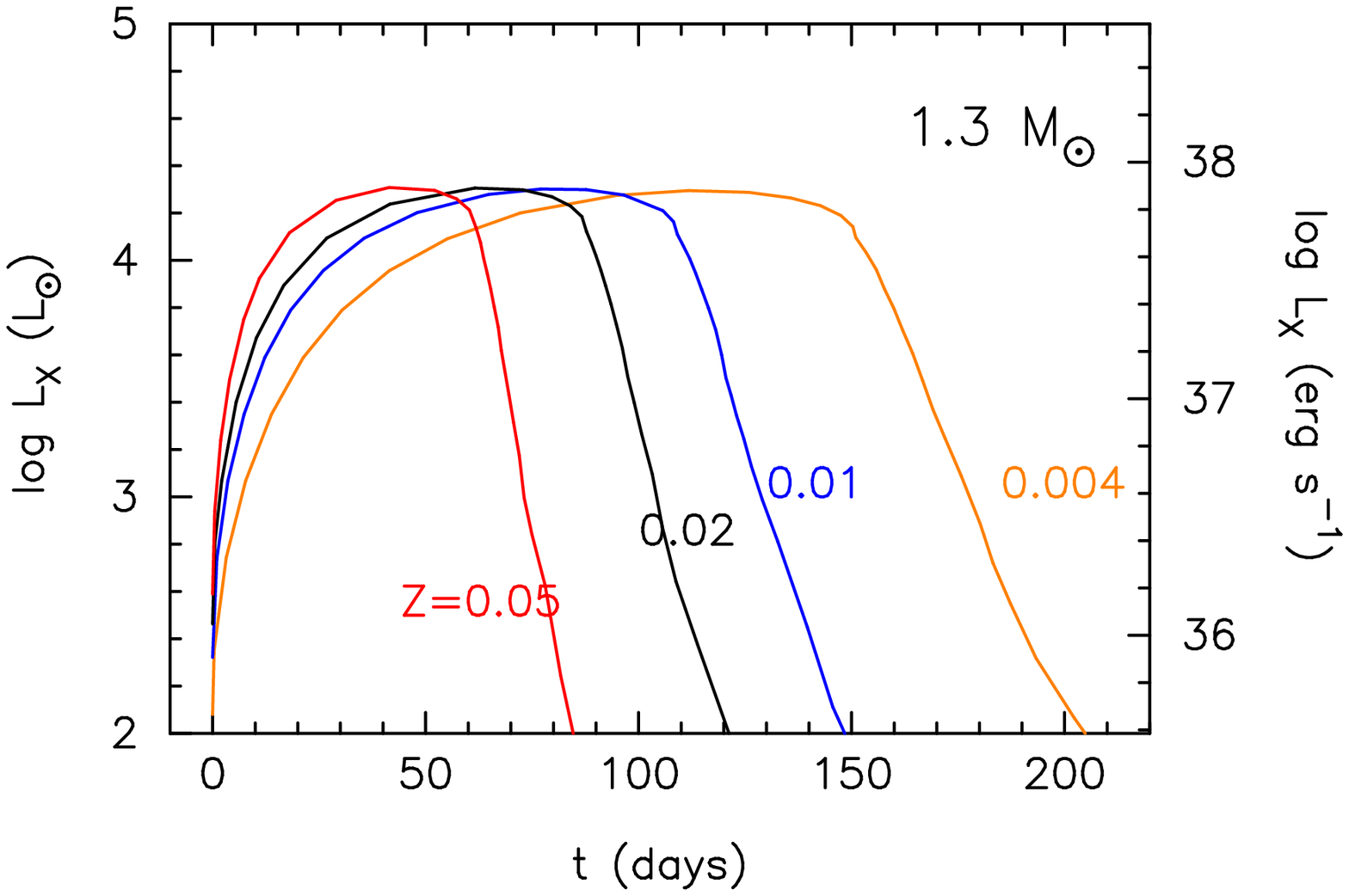}
 \end{center}
\caption{
Same as Figure \ref{sss12}, but for a 1.3 $M_\odot$ WD. 
\label{sss13}}
\end{figure}

\begin{figure}
  \begin{center}
  \includegraphics[width=8.cm]{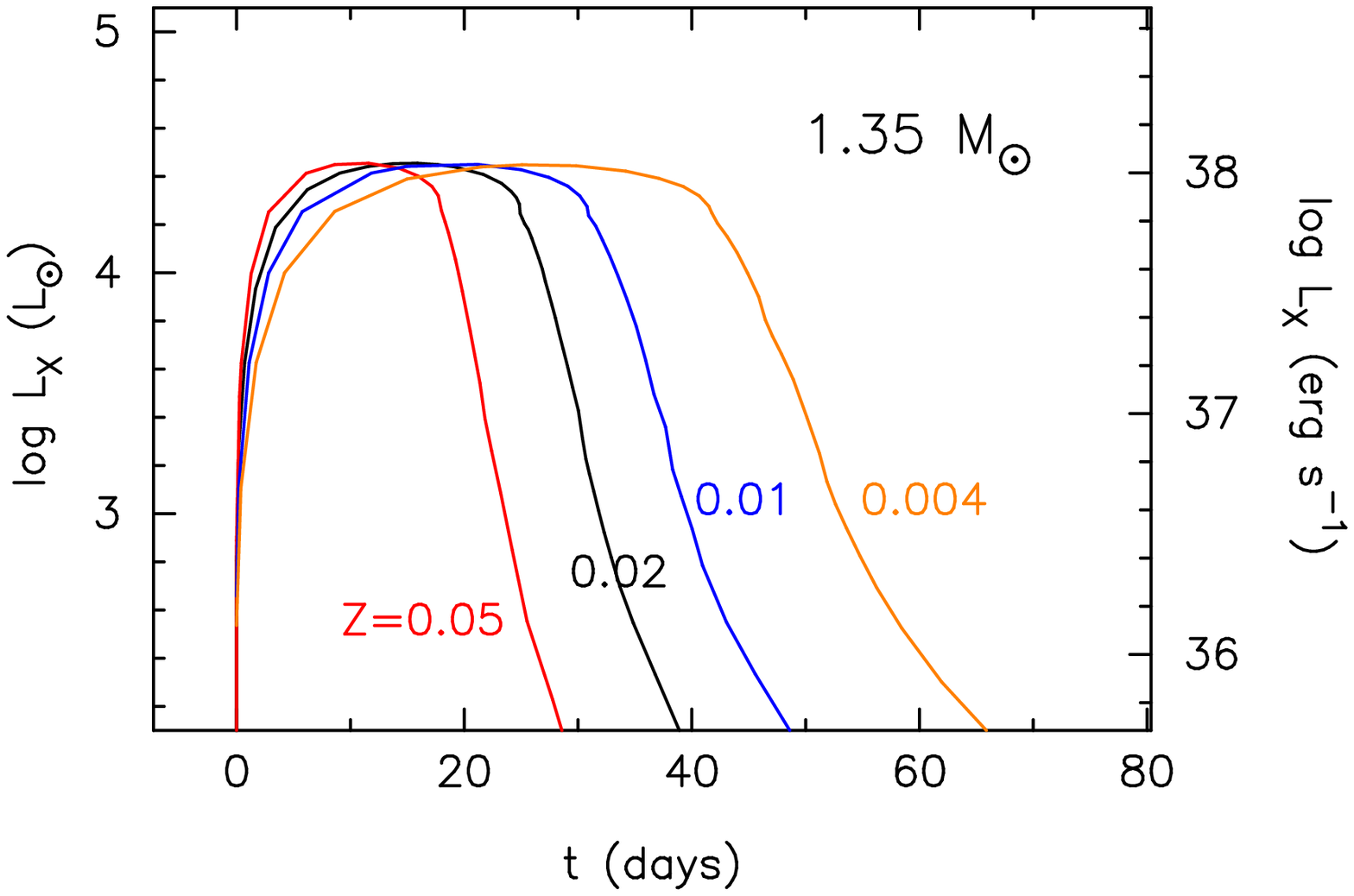}
 \end{center}
\caption{
Same as Figure \ref{sss12}, but for a $1.35~M_\odot$ WD.
\label{sss135}}
\end{figure}

\begin{figure}
  \begin{center}
\includegraphics[width=8.cm]{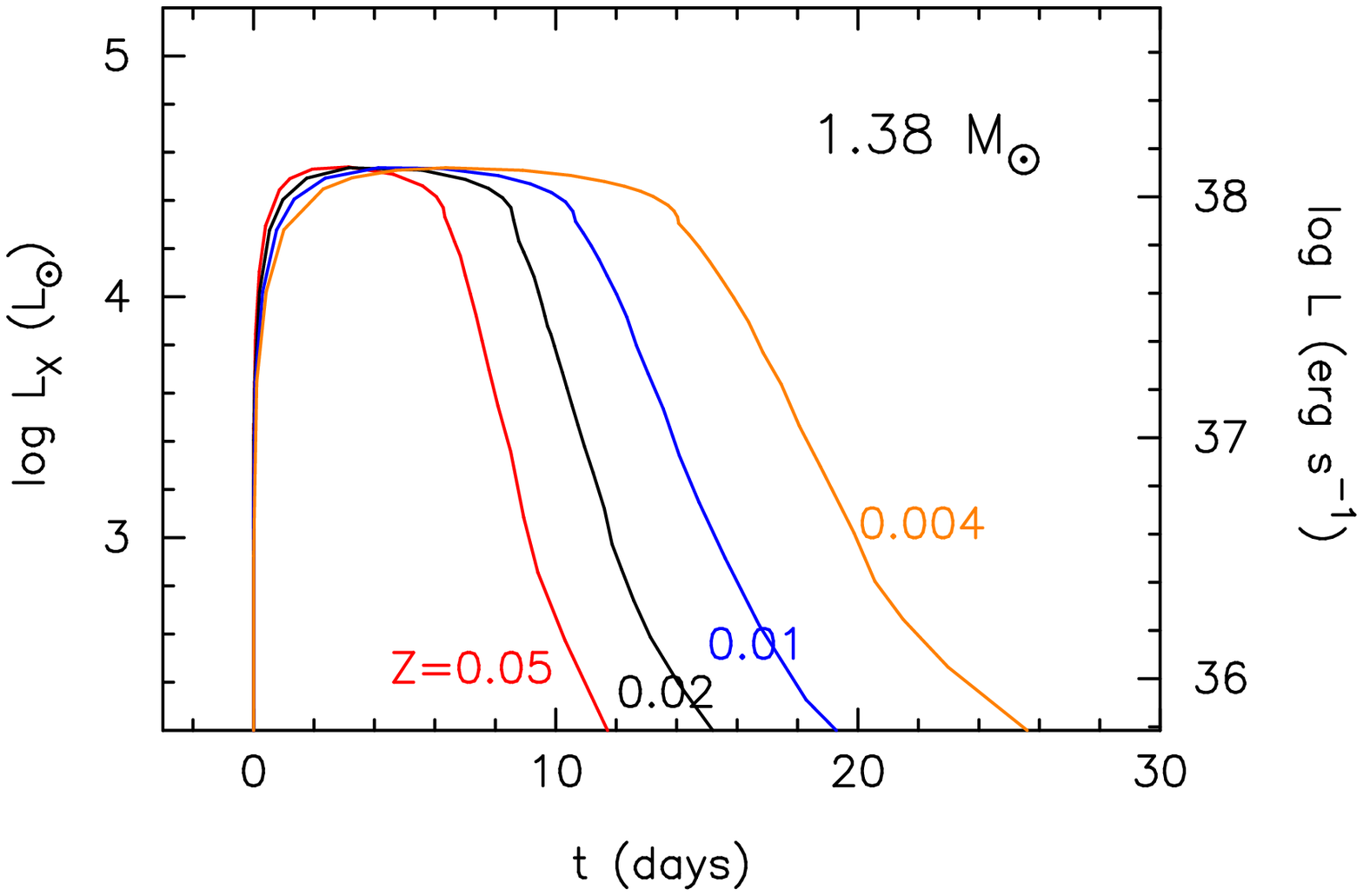}
 \end{center}
\caption{
Same as Figure \ref{sss12}, but for a 1.38 $M_\odot$ WD.
\label{sss138}}
\end{figure}

\section{SSS durations of recurrent novae}
\label{sec_longsss}

\subsection{Numerical Models}

The supersoft X-ray source phase of post nova outburst has been 
theoretically calculated for various WD masses and chemical compositions
of hydrogen-rich envelopes \citep{kat94h, kat99, sal05, wol13, kat13hh}. 
These calculations, however, did not exactly cover the parameter region 
for recurrent novae.  Thus, we have calculated the SSS phases 
and X-ray light curves for extended ranges of parameters. 
The numerical method is the same as that in \citet{kat94h}, in which
the optical decay phase is followed by a sequence of optically-thick
wind solutions. 
After the winds stop, an SSS phase is followed by a sequence of
hydrostatic solutions. 

To calculate the envelope solutions, 
we need to assume the WD radius and chemical composition of the envelope. 
These values are taken from our evolution calculations of shell flashes
already published.  Although the WD accretes solar composition 
($X=0.7$, $Y=0.28$, and $Z=0.02$) material, convection widely develops
at the outburst and a part of nuclear ash helium of the previous outburst
is dredged up and mixed into the whole envelope. As a result, hydrogen mass
fraction decreases by some amount from the original $X=0.7$,
and helium mass fraction increases by the same amount. 
This hydrogen decrease is larger in more massive WDs. 
We adopt WDs of $1.2~M_\odot$, $1.3~M_\odot$, $1.35~M_\odot$,
and $1.38~M_\odot$. The mass accretion rates of our published models do
not cover enough fine grid, but the dependence of composition and radius 
on the mass accretion rate is not large as far as the recurrence period 
is less than a few tens of years.

The envelope chemical composition and WD radius for the $1.2~M_\odot$ WD
are assumed to be $(X,Y,Z)=(0.6,0.38, 0.02)$ and 
$\log (R_{\rm WD}/R_\odot)=-2.200$. 
These values are taken from the time-dependent nova model of 
a $1.2 ~M_\odot$ WD with the mass accretion rate to the WD of
$\dot M_{\rm acc} = 2 \times 10^{-7}~M_\odot$~yr$^{-1}$, 
corresponding to $P_{\rm rec}=9.9$ yr
(at stage G in \citet{kat17sha}).
Here, we assumed the WD radius to be that of the maximum nuclear burning rate 
$\epsilon_{\rm nuc}^{\rm max}$.  
For the 1.3~$M_\odot$ WD, we adopt $(X,Y,Z)=(0.59,0.39, 0.02)$ 
and $\log (R_{\rm WD}/R_\odot)=-2.348$ taken from the model of 
$\dot M_{\rm acc} = 1.0 \times 10^{-7}~M_\odot$~yr$^{-1}$ 
($P_{\rm rec}=10$ yr). 
For the 1.35~$M_\odot$ WD, we adopt $(X,Y,Z)=(0.56,0.42, 0.02)$ and 
$\log (R_{\rm WD}/R_\odot)=-2.462$
from the model of $\dot M_{\rm acc} = 5 \times 10^{-8}~M_\odot$~yr$^{-1}$ 
($P_{\rm rec}=12$ yr)
 \citep{kat16xflash}.
For the 1.38~$M_\odot$ WD, we adopt
$(X,Y,Z)=(0.55,0.43, 0.02)$ and $\log (R_{\rm WD}/R_\odot)=-2.566$
from the model of $\dot M_{\rm acc} = 1.6 \times 10^{-7}~M_\odot$~yr$^{-1}$ 
($P_{\rm rec}=0.95$ yr) 
\citep{kat17sha}.
We have also calculated various population novae, 
i.e., for $Z=0.05$, 0.01, and 0.004. 
Here, we assume the same $X$ for each WD mass and $Y=1.0-X-Z$.  
The X-ray light curve is calculated from 
the photospheric luminosity and temperature of blackbody emission.

Figure \ref{sss12} shows the X-ray ($0.3 - 1.0$ keV) light curve  
for the $1.2~M_\odot$ WD model with different heavy element compositions. 
The black line indicates the $Z=0.02$ model and the dotted line does the 
same WD mass and composition but of the evolution calculation 
with a Henyey type code in \citet{kat17sha}. 
Our static-sequence model (solid black line) shows a good agreement with
the evolution model (dotted black line). 
This figure also shows the models with different metallicities, 
$Z=0.05$, 0.01, and 0.004. 
The SSS duration lasts longer for a smaller $Z$ because the envelope mass 
is larger for a smaller $Z$ at the start of the SSS phase, 
so the nuclear burning lasts longer.   
In an envelope in hydrostatic balance having a smaller $Z$, the envelope mass 
is larger to keep higher temperature in the nuclear burning 
region. Thus, for a smaller $Z$, the envelope mass is larger.

Figures \ref{sss13}, \ref{sss135} and \ref{sss138} show the X-ray
light curves for the 1.3~$M_\odot$, 1.35~$M_\odot$ and 1.38~$M_\odot$ WDs,
respectively.  For a more massive WD, the SSS phase is shorter 
because the envelope mass is small. 
To compare our model light curves with the observations of recurrent novae, 
we define the SSS duration to be the period when the X-ray flux is larger 
than $10 \%$ of its peak, i.e., $L_{\rm X} > 0.1 ~L_{\rm X}^{\rm peak}$.
These theoretical SSS durations are shown in Figure \ref{xonoff.nocorr.rn}(b) 
for $Z=0.02$ by the short horizontal bars. 
The SSS period is shorter for a more massive WD.
The $\Delta (\log ~t_{\rm SSS})$-separation between the $1.3 ~M_\odot$ and 
$1.35 ~M_\odot$ is almost the same as that between the 
$1.35~M_\odot$ and $1.38~M_\odot$. 
It is because the WD radius decreases faster when the WD mass 
approaches the Chandrasekhar mass limit.  This property
makes the envelope mass smaller and then the SSS period 
becomes shorter as the WD mass approaches the Chandrasekhar mass limit.

Figure \ref{xonoff.rn}(b) shows the SSS periods for specified WD masses
with a set of colorful horizontal bars of different $Z$'s. 
The SSS period is shorter for a larger $Z$ as already shown 
in the previous subsection.  The $\Delta (\log ~t_{\rm SSS})$-separation
between two different $Z$, e.g, $Z=0.004$ and $Z=0.02$,
is smaller than that between the two different WD masses, 
e.g., 1.35~$M_\odot$ and 1.38~$M_\odot$. 
Thus, we can estimate the WD mass in the grid of 1.2~$M_\odot$, 
1.3~$M_\odot$, 1.35~$M_\odot$, and 1.38~$M_\odot$ for a recurrent nova 
even if heavy element content $Z$ is poorly known.

\begin{figure*}
\begin{center}
\includegraphics[width=14cm]{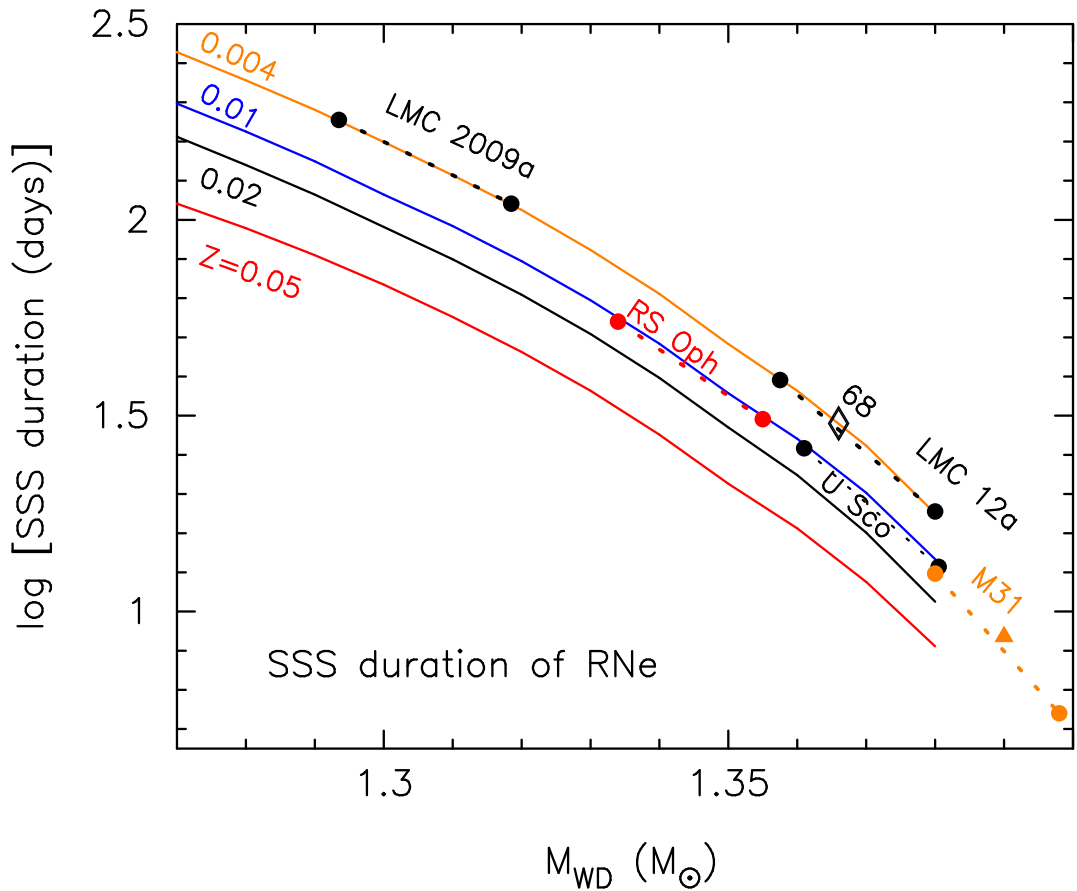}
\end{center}
\caption{
The calculated SSS durations of recurrent novae with various WD masses 
and heavy element contents of $Z=0.004$ (solid orange line), 
0.01 (blue), 0.02 (black), and 0.05 (red). 
The dotted lines with large dots at the both sides 
indicate the SSS duration of individual nova, the name of which
is beside the line.  The left-side dot corresponds to the full width
of SSS duration, $t_{\rm SSS}$, and the right-side dot the SSS duration
after its highly variable phase ended, $t_{\rm SSS}^*$. The filled orange
triangle indicates the 2016 outburst (shorter SSS) of M31N 2008-12a.
The open black diamond shows $t_{\rm SSS}$ of LMC N 1968.   
\label{rn.sssWDmass}}
\end{figure*}

\subsection{Comparison with observed SSS durations}\label{sec_comparison}

Figure \ref{rn.sssWDmass} summarizes the SSS durations calculated 
for various WD masses and metallicities.  
As explained in the previous section, the SSS duration quickly decreases 
as the WD mass increases toward the Chandrasekhar mass.

We also plot the six recurrent novae of a broad rectangular X-ray light curve.
We place the galactic and M31 novae, i.e., U Sco, RS Oph and M31N 2008-12a, 
between the lines of $Z=0.01$ and $Z=0.02$, considering the 
recent estimates for the solar abundance of $Z=0.0138$ \citep{gre19}. 
For the three LMC novae, we adopt $Z=0.004$ as a typical value for the LMC
stars. 

These novae show high variability in their early X-ray phase. 
This variable phase has not been theoretically explained, so we do not know  
whether this phase corresponds to the end of the wind phase or post-wind phase 
in the theoretical models. Thus, we plot both the two durations, 
$t_{\rm SSS}$ and $t_{\rm SSS}^*$, at the left-edge dot and right-edge dot
of each dotted line, respectively. 
The two single symbols represent LMC N 1968 (open black diamond) 
and M31N 2008-12a 2016 outburst (filled orange triangle) 
that have no corresponding $t_{\rm SSS}^*$.

The WD masses in these recurrent novae have been estimated
by various methods. 
For RS Oph, the WD mass is estimated to be 1.35 $M_\odot$ from the optical 
light curve fitting with a composite model of the WD photosphere, 
irradiated accretion disk and companion \citep{hac06b}, 
and also to be 1.35 $M_\odot$ from fitting of the X-ray light curve
after the highly variable phase ended \citep{hac07kl}. 
This value is consistent with our right-side point of the dotted line
for RS Oph.  This suggests that highly variable SSS phase is associated 
with the late wind phase rather than the phase after the wind stops. 
For U Sco, the WD mass is estimated to be 
$M_{\rm WD}=1.55\pm 0.24 ~M_\odot$ by \citet{tho01} 
from the double line orbital velocities,  
and about 1.37 $M_\odot$ by \citet{hac00a} from the light curve fitting. 
These values are consistent with our dotted line. 
For M31N 2008-12a, a 1.38 $M_\odot$ WD is suggested from the
theoretical models \citep{kat15sh,kat16xflash,kat17sha}. 
This value drops on the line 
for M31N 2008-12a in Figure \ref{rn.sssWDmass}. 

For LMC novae, we fitted our SSS periods on the line of $Z=0.004$ in 
Figure \ref{rn.sssWDmass}. 
\citet{hac18b} compared the X-ray and $V$ light curves of LMC N 2009a 
with theoretical light curve of $1.25~M_\odot$ with a neon rich composition 
(Ne nova 3 with $Z=0.02$:  Figure 33a in \citet{hac18b}).  
Note that this is a tentative fitting because no neon-enrichment 
is observed in LMC 2009a. Moreover,  
LMC N 2009a is an LMC member and we expect a lower metallicity ($Z < 0.02$)
and no substantial enrichment of neon.
A nova evolution becomes shorter for a higher metallicity 
\citep{kat97,kat99}, thus, we expect a more massive WD than 1.25 $M_\odot$ 
for a lower metal environment. 
Thus, our value of $1.29-1.32 ~M_\odot$ in Figure \ref{rn.sssWDmass}
is reasonable.  
For LMC N 2012a, \citet{hac18b} estimated the mass to be $1.37~M_\odot$ 
from a linear interpolation of the stretching factor. 
This value is consistent with our value in Figure \ref{rn.sssWDmass}. 

In this way, our WD masses obtained from the SSS duration 
are consistent with the other theoretical and observational estimates. 
We conclude that the SSS duration of a recurrent nova
is a good indicator of the WD mass 
if the X-ray light curve shows a broad rectangular shape.

\section{Discussion} \label{sec_discussion}

\subsection{RS Oph and V2491 Cyg}\label{sec_v2491cyg}

\citet{hac19a} compared the light curves of RS Oph and V2491 Cyg, 
and showed that $t_{\rm on}$ is almost the same, while $t_{\rm off}$
is very different (see their Figure 16). 
Figure \ref{xonoff.nocorr.rn} clarifies that this difference can 
be understood along with the increasing deviation of recurrent novae 
from classical novae toward a more massive WD.
The WD mass of RS Oph has been estimated to be $1.35 ~M_\odot$ 
(see Section \ref{sec_longsss}). 
Also the similar value, $M_{\rm WD} \sim 1.35 ~M_\odot$, is obtained in 
V2491 Cyg \citep{hac19a} from the light curve fitting. 
In short, the WD mass is similar but the SSS duration is very different. 

There are several reasons that result in a longer SSS period 
in recurrent novae 
than in classical novae having the same WD mass.  
The first is a longer variable period in the SSS phase. 
Both RS Oph and V2491 Cyg show a highly variable SSS phase, 
but RS Oph shows a much longer variable period. 
The second is the difference in the chemical composition of envelope. 
The ejecta is close to the solar composition in recurrent novae
whereas the ejecta is enriched with heavy elements and 
deficient in hydrogen in classical novae (e.g., Table 1 in \citet{hac06k}). 
A smaller $X$ provides less nuclear fuel that results in a shorter 
nuclear-burning time. Also, the heavy element enrichment causes 
a smaller envelope mass. Thus, the SSS period is shorter. 
The last reason is high mass-accretion rates in recurrent novae that 
release larger gravitational energy per unit time.  
The WD core is heated and slightly expands.
This makes the envelope mass larger in hydrostatic balance of SSS phase,
and therefore the SSS phase becomes longer. 
To summarize, the SSS phases in recurrent novae are systematically
longer than those in classical novae having the same WD mass  
because of the combination of these several effects.

For V2491 Cyg, \citet{mun11} obtained the composition
of ejecta to be $X = 0.573$, $Y = 0.287$, and $Z = 0.140$  
by weight. 
Here, the individual elements are $X_{\rm N}= 0.074$, 
$X_{\rm O} = 0.049$, and $X_{\rm Ne} = 0.015$. 
This heavy element enrichment strongly suggests that V2491 Cyg is not
a recurrent nova but a classical nova, because such large amounts of
heavy elements have not been detected in recurrent novae.
Comparing with our model for RS Oph in Section \ref{sec_longsss}, 
$X=0.56$ and $Z=0.02$ for the $1.35 ~M_\odot$, 
the hydrogen content is almost the same, but the
heavy elements are much larger in V2491 Cyg. 
Even if the WD masses and $X$'s are the same between RS~Oph and V2491~Cyg,
the envelope mass when the winds stop is smaller for larger $Z$ 
(heavy element enrichment) and, as a result, the SSS phase of V2491~Cyg
is much shorter.

\subsection{T Pyx}\label{sec_tpyx}

T Pyx also shows a shorter SSS phase than the trend of recurrent novae 
as shown in Figure \ref{xonoff.nocorr.rn}. 
\citet{sch10} examined the radial motion of ejecta knots around T Pyx 
and the quiescent $B$ magnitude decay since 1890 to 2009. 
They concluded that T Pyx experienced a first classical nova outburst 
around 1866 after a long period of very low mass-accretion rates, and it  
is currently staying in the period of very high mass-accretion rates, as 
indicated by several successive recurrent nova outbursts. 
In short, the mass accretion rate has long been very low 
before 1866, but increased after that. 
In such a case the WD interior is still cool, has not yet become hot. 
So the WD radius is possibly smaller than that of the hot WD 
corresponding to the high mass-accretion rates. 
Thus, the hydrogen-rich envelope mass is smaller than those of 
recurrent novae with the same WD mass. 
This is a possible explanation of the short SSS phase of T Pyx.  
We also point out that the duration of the SSS phase is close to
the line for classical novae (blue line in Figure
\ref{xonoff.nocorr.rn}).

\begin{figure*}
\begin{center}
\includegraphics[width=14cm]{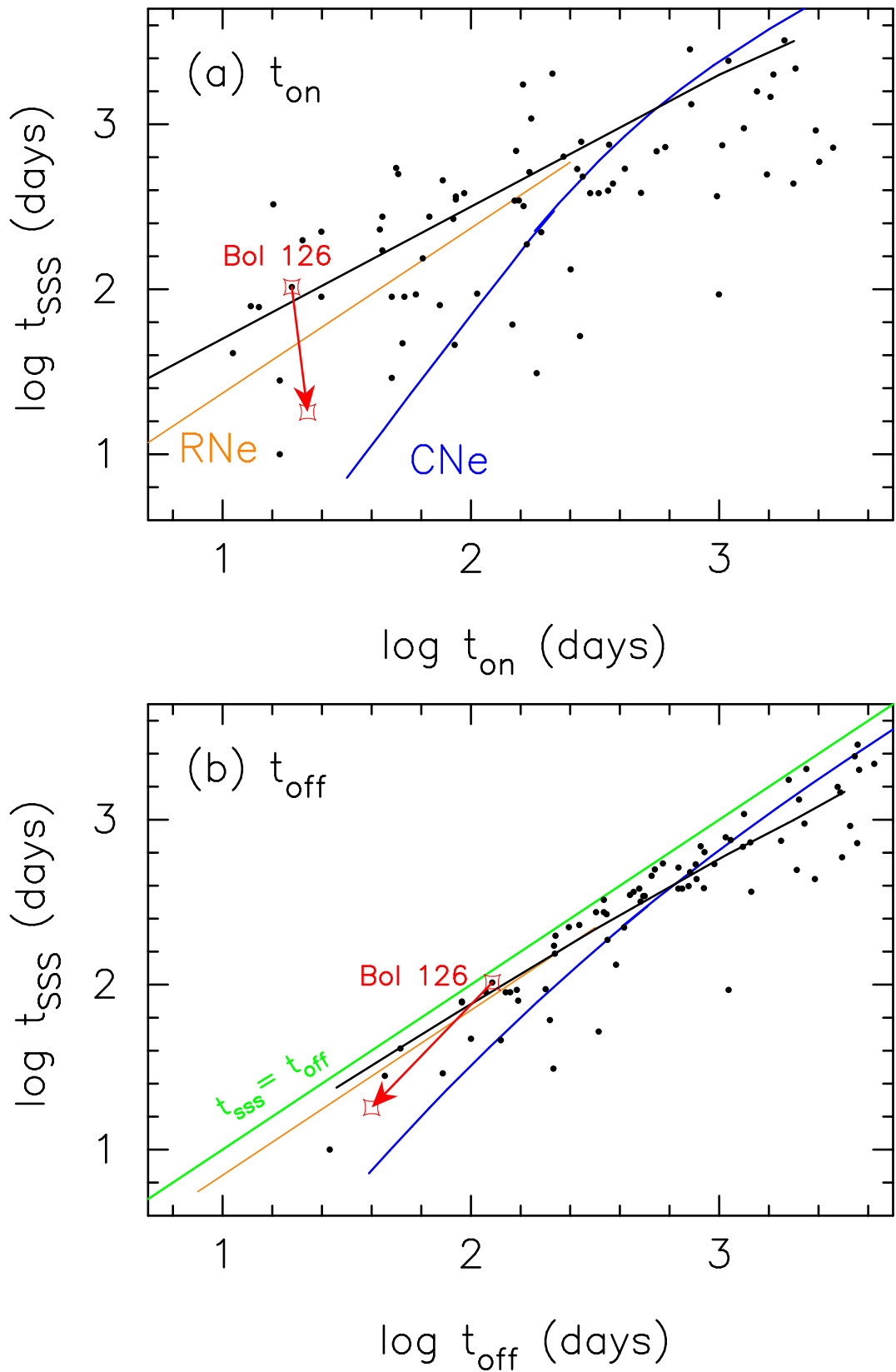}
\end{center}
\caption{
Comparison of the distribution of M31 novae with our empirical laws. 
The M31 novae are plotted with the black dots (data taken from \citet{hen14}). 
The solid black line indicates a relation made from the linear fit in the 
$(\log~t_{\rm on})$-$(\log~t_{\rm off})$ diagram by \citet{hen14}. 
The other lines are the same as those in Figure \ref{xonoff.nocorr.rn}, i.e.,
recurrent novae (orange), classical novae (blue), and
$t_{\rm SSS}=t_{\rm off}$ (green).
The red curved squares connected by the arrow indicate
M31N 2010-10f; the upper square is by \citet{hen14} and the lower symbol
is the position under our definition. 
\label{rncn}}
\end{figure*}

\subsection{Comparison with M31 Classical Novae}\label{sec_m31n}

In the present work, we concentrate on the recurrent novae 
and have not discussed much on classical novae. 
Here, we compare our results with the statistical work 
for classical novae appeared in M31 \citep{hen11,hen14}. 

Figure 8(a) in \citet{hen14} shows the $t_{\rm on}$ - $t_{\rm off}$ diagram 
in which the M31 classical novae distribute around a line of 
Equation (\ref{equ_henze}) with a relatively large scatter. 
This large scatter may be partly because of the sparse observation 
with telescopes and X-ray satellites that causes large 
ambiguities in the outburst day $t_0$, $t_{\rm on}$, and $t_{\rm off}$.

From Equation (\ref{equ_henze}) we made 
a $t_{\rm SSS}$ - $t_{\rm off}$ relation and depicted it by  
the black line in Figure \ref{rncn}, as well as the SSS data 
for the 74 classical novae in M31 (black dots) taken from \citet{hen14}. 
The black line is close to our line of recurrent novae rather than  
classical novae. The individual nova points, however, distribute 
widely and covers all the three lines. 

There is a reason that we cannot directly compare these data 
with our results. 
It is the difference in the definition of 
$t_{\rm on}$ and $t_{\rm off}$. 
\citet{hen14} obtained the X-ray turn-on time, $t_{\rm on}$, 
as the middle of the last non-detection and the first detection time. 
Similarly, the turnoff time $t_{\rm off}$ is obtained as the mean 
of the last detection and non-detection times.
On the other hand, our definition of turn on/off time is 
the time when the X-ray flux reaches one tenth of the peak flux. 
As a result, our $t_{\rm SSS}$ is likely to be shorter than Henze et al.'s 
if we have relatively dense observation. 

For example, LMC N 2009a (Figure \ref{lmc2009}) is a well-observed 
nova in which the faint X-ray flux ($\sim$ one thousandth of the peak flux)
is detected before and after the bright SSS phase. If we include these 
faint phase, we get an earlier $t_{\rm on}$, later $t_{\rm off}$, and
much longer $t_{\rm SSS}$.

More exactly, we see the case of M31N 2010-10f in Figure \ref{rncn}. 
The black dot outlined by a curved red square 
represents the data in \citet{hen14}, while 
the lower square is the one measured with our definition 
from the X-ray light curve presented by \citet{hen13} and \citet{kat13hh}.  
Henze et al.'s value is larger than ours by a factor of 5.7, 
because they counted the post outburst faint X-rays. 

In this way, it is hard to directly compare our results with the statistical 
data by \citet{hen14}. 
To draw a qualitative conclusion we need to adopt the same 
definition of the X-ray phase.

\section{Conclusions}\label{sec_conclusion}

We examined supersoft X-ray light curves of seven recurrent novae 
and one candidate. 
Our main results are summarized as follows. 

\begin{enumerate}
\item
Six novae out of eight, M31N 2008-12a, LMC N 1968, U Sco, LMC N 2012a, 
RS Oph, and LMC N 2009a, show a broad rectangular X-ray light curve shape. 
The X-ray phase is highly variable in the first half period. 
The optical light curve shows a plateau during the SSS phase
that indicates the presence of a large irradiated accretion disk. 

\item 
These six novae show a common proportionality relation 
between the SSS duration and total nova duration, 
$t_{\rm SSS}=0.70 ~t_{\rm off}$. If we exclude the highly
variable phase from the SSS phase, the relation is 
$\log ~t_{\rm SSS}^*=1.08 ~ (\log ~t_{\rm off}-1.2)+ 0.7$. 

\item 
The other two novae, V745 Sco and T Pyx, show a narrow triangular
shape of X-ray light curve without a highly-variable phase. 
Their SSS phases are shorter than the proportionality relation of
$t_{\rm SSS}=0.70 ~ t_{\rm off}$.  
The $t_{\rm SSS}$ of V745~Sco suddenly drops from the line of
$t_{\rm SSS}=0.70 ~ t_{\rm off}$ at the shortest limit of $t_{\rm off}$.
The $t_{\rm SSS}$ of T~Pyx is slightly below the line of
$t_{\rm SSS}=0.70 ~ t_{\rm off}$ at the longer edge of $t_{\rm off}$
for recurrent novae, and rather close to the $t_{\rm SSS}$-$t_{\rm off}$
relation for classical novae proposed by \citet{hac10k}. 
The resemblance to classical novae is consistent with the observational
suggestion that T~Pyx has recently become a recurrent nova
in 1866 and had undergone a very long quiescent phase before that.  

\item 
The 2016 outburst of M31N 2008-12a shows a shorter $t_{\rm SSS}$ and
$t_{\rm off}$ than those of the other year's, the position of which is
located in the middle of usual M31N 2008-12a and V745~Sco.
We consider the 2016 outburst (shorter duration) to be the key to 
examine whether V745 Sco is the shortest end of the bending line
near the edge or merely an exception to the proportionality relation.

\item 
We present theoretical SSS durations calculated for recurrent novae 
with various WD masses and stellar metallicities 
($Z=$0.004, 0.01, 0.02 and 0.05).  
Comparing the observed SSS phase with the model calculation, 
we estimate the WD mass for the five recurrent novae with a broad
rectangular SSS shape. 
These WD masses are in good agreement with those estimated 
from the other methods.  
The duration of the SSS phase of a recurrent nova is a good indicator
of the WD mass. 

\end{enumerate}


\begin{ack}
We are grateful to Martin Henze for providing us the M31 nova data.
We also thank the anonymous referee for useful comments that
improved the manuscript.
\end{ack}

\appendix 

\begin{figure}
 \begin{center}
  \includegraphics[width=9cm]{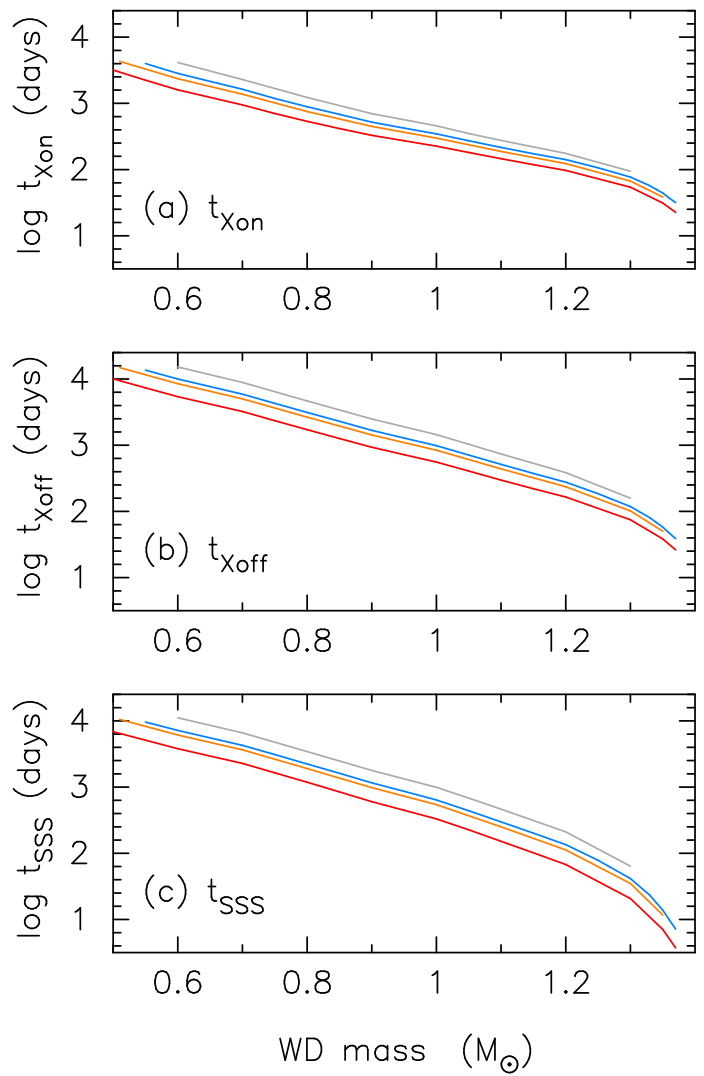}
 \end{center}
\caption{
(a) The X-ray turn-on time, (b) X-ray turnoff time, and 
(c) duration of the supersoft X-ray phase of classical novae 
against the WD mass. 
The different color indicates the different chemical composition of 
typical classical novae \citep{hac10k}.
From upper to lower. 
Gray line: Ne nova 3 
($X,~Y,~X_{\rm CNO},~X_{\rm Ne})$=(0.65, 0.27, 0.03, 0.03). 
Blue line: Ne nova 2 (0.55, 0.3, 0.1, 0.03).
Orange line: CO nova 4 (0.55, 0.23, 0.2, 0.0). 
Red line: CO nova 3 (0.45, 0.18, 0.35, 0.0). 
For all the cases the heavy element abundance is assumed to be $Z=0.02$
and $X+Y+X_{\rm CNO}+X_{\rm Ne}+Z=1$.    
}\label{wdmass.sss}
\end{figure}

\begin{figure}
 \begin{center}
  \includegraphics[width=8cm]{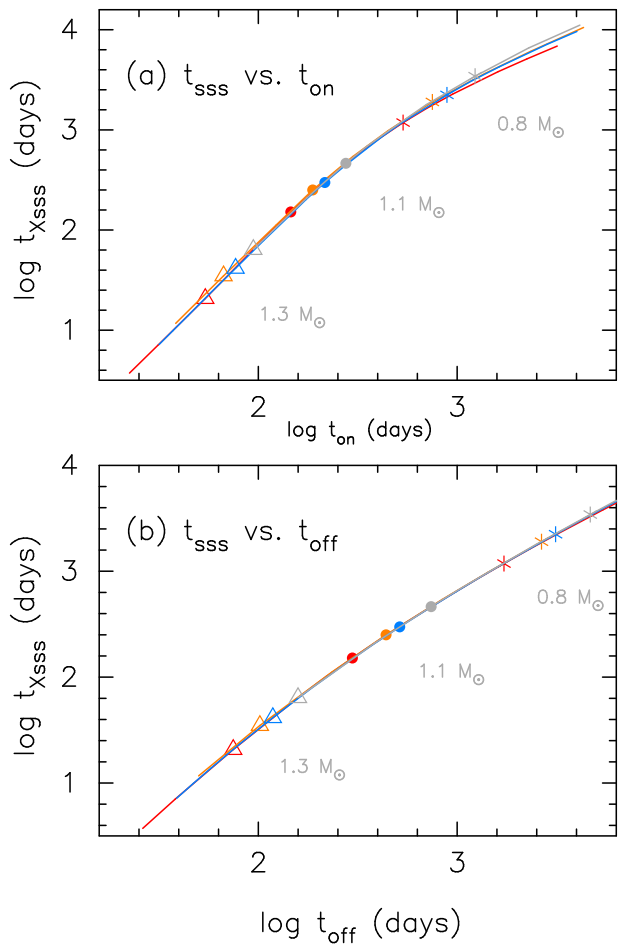}
 \end{center}
\caption{
Duration of the supersoft X-ray phase of classical novae against 
(a) X-ray turn-on time, and (b) X-ray turnoff time. 
The color corresponds to the model of the same color 
in Figure \ref{wdmass.sss}. 
The positions of the three WD masses are indicated on each line 
with the same color as the line. 
Open triangles: 1.3~$M_\odot$.  Dots: 1.1~$M_\odot$.
Asterisks: 0.8~$M_\odot$. 
}\label{xonoffWDmass}
\end{figure}

\section{Theoretical relation of $t_{\rm SSS}-t_{\rm off}$ for classical novae}
\label{tsss-versus-toff}

The decay phase of a nova can be followed by a sequence of 
steady-state and static envelope solutions \citep{kat94h}.  
\citet{hac10k} calculated $t_{\rm on}$ and $t_{\rm off}$ 
for various WD masses and chemical compositions. 
The SSS duration is calculated by $t_{\rm SSS}=t_{\rm off}-t_{\rm on}$.   
Figure \ref{wdmass.sss} shows $t_{\rm on}$, $t_{\rm off}$, and $t_{\rm SSS}$ 
for various WD masses and different chemical compositions of 
hydrogen-rich envelopes. 
The four lines represent typical compositions of novae \citep{hac10k};  
Ne nova 3 ($X,~Y,~X_{\rm CNO},~X_{\rm Ne})= (0.65, 0.27, 0.03, 0.03)$ 
(gray line), 
Ne nova 2 $(0.55, 0.3, 0.1, 0.03)$ (blue line), 
CO nova 4 $(0.55, 0.23, 0.2, 0.0)$ (orange line), and 
CO nova 3 $(0.45, 0.18, 0.35, 0.0)$ (red line).
For all the cases, we assume $Z=0.02$ and $X+Y+Z+X_{\rm CNO}+X_{\rm Ne}=1$.

\citet{kat94h,hac06k,hac10k} clarified that the nova evolution speed
depends strongly on the WD mass and weakly on the chemical composition.
As in Figure \ref{wdmass.sss}, a nova on a more massive WD evolves faster 
so both the $t_{\rm on}$ and $t_{\rm off}$ times are smaller.  
For the same WD mass, the nova evolution is faster for smaller $X$ and for 
larger $X_{\rm CNO}$. 
Thus, both the $t_{\rm on}$ and $t_{\rm off}$ are smaller.

Figure \ref{xonoffWDmass} shows the SSS duration of the same models 
as in Figure \ref{wdmass.sss}, but against the X-ray (a) turn-on and
(b) turnoff time.  All the four lines converge into one. 
Although the lines are almost converged, the WD mass associated 
with each line is different. 
The points corresponding to the $0.8~M_\odot$, $1.1~M_\odot$
and $1.35 ~M_\odot$ are indicated on each line. For the same WD mass,
the SSS duration is longer for a larger hydrogen content $X$.

Figures \ref{xonoff.nocorr.rn} and \ref{xonoff.rn} show the relations
between $t_{\rm on}$, $t_{\rm off}$, and  $t_{\rm SSS}$ 
taken from Figure \ref{xonoffWDmass}. 
The line is a combination of CO nova 3 ($\leq 1.05~M_\odot$) and Ne nova 2
($\geq 1.05~M_\odot$) as a representative of typical novae 
(Figure 22 in \citet{hac19b}). 
These two lines are smoothly connected with a small overlap around 
$\log ~t_{\rm SSS}$ (days) =2.4. 
Even if we adopt the other chemical composition,
this line hardly changes as shown in Figure \ref{xonoffWDmass}.

%


\begin{thebibliography}{}


\bibitem[Bode et al. (2016)]{bod16}
Bode, M. F., Darnley, M. J., Beardmore, A. P., et al. 2016, \apj, 818, 145


\bibitem[Chomiuk et al. (2014)]{cho14}
Chomiuk, L, Nelson, T., Mukai, K. et al. 2014, \apj, 788, 130


\bibitem[Darnley et al. (2016)]{dar16} Darnley, M. J., et al. 2016, \apj, 833, 149

\bibitem[Darnley et al. (2015)]{dar15} Darnley, M. J., et al. 2015, \aap, 580, A45

\bibitem[Drake et al. (2016)]{dra16}
Drake, J., J., Delgado, L., Laming, J. M. et al. 2016, \apj, 825, 95 


\bibitem[Evans et al. (2009)]{eva09}
Evans, P. A., Beardmore, A. P., Page, K. L., et al.  2009, \mnras, 397, 1177

\bibitem[Grevesse (2019)]{gre19}
Grevesse, N. 2019, 
Bulletin de la Soci\'et\'e Royale des Siences de Li\'ege, 88, 5



\bibitem[Hachisu \& Kato (2000a)]{hac00a} Hachisu, I., \& Kato, M. 2000, \apj, 528, L97

\bibitem[Hachisu \& Kato (2000b)]{hac00b} Hachisu, I., \& Kato, M. 2000, \apj, 534, L189



\bibitem[Hachisu \& Kato (2006)]{hac06k}
Hachisu, I., \& Kato, M. 2006, \apjs, 167, 59

\bibitem[Hachisu \& Kato (2007)]{hac07k}
Hachisu, I., \& Kato, M. 2007, \apj, 662, 552

\bibitem[Hachisu \& Kato (2009)]{hac09} Hachisu, I., \& Kato, M. 2009, \apj, 694, L103

\bibitem[Hachisu \& Kato (2010)]{hac10k}
Hachisu, I., \& Kato, M. 2010, \apj, 709, 680

\bibitem[Hachisu \& Kato (2015)]{hac15k}
Hachisu, I., \& Kato, M. 2015, \apj, 798, 76

\bibitem[Hachisu \& Kato (2016a)]{hac16ka}
Hachisu, I., \& Kato, M. 2016, \apj, 816, 26

\bibitem[Hachisu \& Kato (2016b)]{hac16kb}
Hachisu, I., \& Kato, M. 2016, \apj, 824, 22

\bibitem[Hachisu \& Kato (2016c)]{hac16c}
Hachisu, I., \& Kato, M. 2016, \apjs, 223, 21

\bibitem[Hachisu \& Kato (2018a)]{hac18a}
Hachisu, I., \& Kato, M. 2018a, \apj, 858, 108

\bibitem[Hachisu \& Kato (2018b)]{hac18b}
Hachisu, I., \& Kato, M. 2018b, \apjs, 237, 4 

\bibitem[Hachisu \& Kato (2019a)]{hac19a}
Hachisu, I., \& Kato, M. 2019a, \apjs, 241, 4 

\bibitem[Hachisu \& Kato (2019b)]{hac19b}
Hachisu, I., \& Kato, M. 2019b, \apjs, 242, 18

\bibitem[Hachisu et al. (2006b)]{hac06b}
Hachisu, I., Kato, M., Kiyota, S., et al. 2006b, \apjl, 651, L141

\bibitem[Hachisu et al. (2007)]{hac07kl}
Hachisu, I., Kato, M., \& Luna, G. J. M. 2007, \apj, 659, L153








\bibitem[Hachisu et al. (2016)]{hac16sk}
Hachisu, I., Saio, H., \& Kato, M.  2016, \apj, 824, 22 

 
\bibitem[Henze et al. (2015)]{hen15}
Henze, M., Ness, J.-U., Darnley, M., et al. 2015, \aap, 580, 46 

\bibitem[Henze et al. (2018)]{hen18}
Henze, M., Darnley, M., Williams, S. C. et al. 2018, \apj, 857, 68

\bibitem[Henze et al. (2013)]{hen13} Henze, M., Pietsch, W., Haberl, F., et al. 2013, 549, 120


\bibitem[Henze et al. (2014)]{hen14} Henze, M., Pietsch, W., Habert, F., 
Della Valle, M., Sala, G,  Hatzidimitriou, D., Hofmann, F., Hernanz, M. et al. 2014, \aap, 563, A2

\bibitem[Henze et al. (2011)]{hen11} Henze, M., Pietsch, W., Haberl, F., 
Hernanz, M., Sala, G., Hatzidimitriou, D., Della Valle, M.,  Rau, A. 
et al. 2011, \aap, 533, A52



\bibitem[Kato (1997)]{kat97} Kato, M. 1997, \pasj, 113, 121

\bibitem[Kato (1999)]{kat99} Kato, M. 1999, \pasj, 51, 525


\bibitem[Kato (2011)]{kat11} 
Kato, M. \ 2011, in Binary Paths to Type Ia Supernovae Explosions, 
IAU symposium no. 281, ed.\  R.~Di Stefano, M. Orio, \& M. Moe 
(Cambridge: Cambridge University Press), 172



\bibitem[Kato \& Hachisu (1994)]{kat94h}
Kato, M., \& Hachisu, I. 1994, \apj, 437, 802


\bibitem[Kato \& Hachisu (2012)]{kat12Review}
 Kato, M., \& Hachisu, I. 2012, Bull.Astr.Soc,India, 40, 393

\bibitem[Kato et al. (2013)]{kat13hh} Kato, M., Hachisu, I., \& Henze, M. 2013, \apj, 779, 19 

\bibitem[Kato et al. (2017)]{kat17palermo}
 Kato, M., Hachisu, I.,  \& Saio, H.  2017, in
The Golden Age of Cataclysmic Variables and Related
Objects - IV, ed. F. Giovannelli et al. (Trieste: SISSA PoS), 315, 56


\bibitem[Kato et al. (2014)]{kat14shn}
Kato, M., Saio, H., Hachisu, I., \& Nomoto, K. 2014, \apj, 793, 136

\bibitem[Kato et al. (2015)]{kat15sh}
Kato, M., Saio, H., \& Hachisu, I. 2015, ApJ, 808, 52


\bibitem[Kato et al. (2017a)]{kat17sha}
Kato, M., Saio, H., \& Hachisu, I.  2017a, \apj, 838, 153
 

\bibitem[Kato et al. (2018)]{kat18sh}
 Kato, M., Saio, H., \& Hachisu, I., 2008, \apj, 863, 125

\bibitem[Kato et al. (2016)]{kat16xflash}
 Kato, M., Saio, H., Henze, M. et al. 2016, \apj, 830, 40

\bibitem[Kuin et al. (2020)]{kui20} Kuin, N.P., Page, K.L., Mr\'oz, P., Darnley, M. J., Shore, S. N. et al 2020, \mnras, 491, 655


\bibitem[Munari et al. (1990)]{mun90}
Munari, U., Margoni, R., \& Stagni, R. 1990, \mnras, 242, 653



\bibitem[Munari et al. (2011)]{mun11} 
Munari, U., Siviero, A., and Dallaporta, S. 2011, New Astronomy, 16, 209


\bibitem[Ness et al. (2012)]{nes12}
Ness, J.-U., Schaefer, B. E., Dobrotka, A., et al. 2012, \apj, 745, 43
 

\bibitem[Nelson et al. (2012)]{nel12a}
Nelson, T., Donato, D., Mukai, K., Sokoloski, J., \& Chomiuk, L. 2012,
\apj, 748, 43,


\bibitem[Nelson et al. (2014)]{nel14} 
Nelson, T., Chomiuk, L. Roy, N. et al. (2014) \apj, 785,78

\bibitem[Ness et al. (2011)]{nes11} 
Ness, J.-U., Osborne, J.P., Dobrotka, A., et al. \apj, 733, 70



\bibitem[Orio et al. (2013)]{ori13} 
Orio, M., Behar, E., Gallagher, J. et al. 2013, \mnras, 429, 1342

\bibitem[Orio et al. (2015)]{ori15}
Orio, M., Rana, V., Page, K. L., Sokoloski, J., \& Harrison, F. 2015, \mnras, 448, L35


\bibitem[Osborne et al. (2011)]{osb11} 
Osborne, J. P., Page, K. L., Beardmore, A.P., et al.   
2011, \apj, 727, 124

\bibitem[Page et al. (2010)]{pag10} 
Page, K.L., Osborne, J.P., Evans, P.A. et al., 2010, \mnras, 401, 121

\bibitem[Page et al. (2015)]{pag15} 
Page, K.L., Osborne, J.P., Kuin, N.P.M. et al., 2015, \mnras, 454, 3108

\bibitem[Payne-Gaposchkin (1957)]{pay57}
Payne-Gaposchkin, C. 1957, The galactic Novae (Amsterdam: North-Holland)


\bibitem[Sala \& Hernanz (2005)]{sal05} 
Sala, G., \& Hernanz, M. 2005, \aap, 439, 1061 


\bibitem[Schaefer (2010)]{schaefer10}
Schaefer, B. E. 2010, \apjs, 187, 275

\bibitem[Schaefer et al. (2010)]{sch10} 
Schaefer, B. E., Pagnotta, A., \& Shara, M. M. 2010, \apj, 708 381

\bibitem[Schaefer et al. (2013)]{sch13}
Schaefer, B.,E., Landolt, A.U., Linnolt, M. et al. 2013, \apj, 773, 55


\bibitem[Schwarz et al. (2015)]{schw15}
Schwarz, G. J., Shore, S. N., Page, K. L., et al. 2015, \aj, 149, 95

\bibitem[Strope et al. (2010)]{str10}
Strope, R., Schaefer, B. E., \& Henden, A. A. 2010, \aj, 140, 34

\bibitem[Thoroughgood et al. (2001)]{tho01}
Thoroughgood, T. D., Dhillon, V. S., Littlefair, S. P., Marsh, T. R.,
\& Smith, D. A. 2001, \mnras, 327, 1323



\bibitem[Walter et al. (2012)]{wal12}
Walter, F. M., Battisti, A., Towers, S. E., Bond, H. E.,
\& Stringfellow, G. S. 2012, \pasp, 124, 1057


\bibitem[Wolf et al. (2013)]{wol13}
Wolf, W. M., Bildsten, L., Brooks, J., \& Paxton, B. 2013, \apj, 777, 136, 






\end{thebibliography}
\end{document}